# CLASSIFICATION OF UNITARY HIGHEST WEIGHT REPRESENTATIONS FOR NON COMPACT REAL FORMS

JUAN GARCIA-ESCUDERO

Departamento de Física
Universidad de Oviedo
33007 Oviedo (Spain)

MIGUEL LORENTE

Departamento de Física
Universidad de Oviedo
33007 Oviedo (Spain)

**ABSTRACT**

**Using Jakobsen theorems, unitarizability in Hermitian Symmetric Spaces is discussed. The set of all missing highest weights is explicitly calculated and the construction of their corresponding highest weights vectors is studied.**

## I.- INTRODUCTION

One of the new methods for the construction of representations for semisimple Lie algebras is based on enveloping algebras.

Irreducible representations of simple Lie Algebras arise if we take the quotient space of Verma modules with respect to some invariant subspaces generated by highest weight vectors (Ref [1]). When a scalar product is induced in these Verma modules infinitesimally unitary representations can be defined (Ref [2])

In this paper we discuss the unitarizability on non compact real forms following the Jakobsen method, (Ref [3], [4], [5]). He uses the Bernshtein, Gel'fand and Gel'fand theorem and the scalar product induced by a sesquilinear form introduced by Harish-Chandra in Ref [6] (see also Ref [7]). We use his method to obtain a complete and explicit classification of the highest weights that we must exclude in order to unitarize when the reduction level is strictly higher than one. In the examples and, when the expressions are not extremely long, we illustrate the procedure by writing the corresponding highest weight vectors that generate the invariant subspaces in which vanishes the sesquilinear form.

There exists other methods in the literature (see Ref [8], [9], [10] and [11]) following differents paths but arriving at the same final results.

Indecomposable representations have found application in physics for a long time (see Ref [12]). So certain types of indecomposable representations are associated with the Poincaré algebra, the algebra of the Euclidean group and others (see Ref [13]). For application of the algebras treated here see Ref [14] and [15].

In this work we consider hermitian symmetric spaces for which the reduction level may be higher than one: $\mathrm{su}(p, q)$, $\mathrm{sp}(n,R)$, $\mathrm{so}^*(2n)$, $e_6$ and $e_7$. In Section II we give some concepts which will be needed. In Section III we describe the Jakobsen method by means of a step series and we state how to construct the highest weight vectors. In Section IV we introduce the concept of height and we give a notation which allow us to localize easily the non compact roots in Jakobsen diagrams. In addition we use the Jakobsen diagrams to obtain, in a very simple way, the split rank which is useful for the calculation of the $\lambda_S$ parameter. Finally in Section V we first study general cases giving the sets of all highest weights which will be missing and then apply the method to some examples.

## II.- **PRELIMINARIES**

Let $g$ be a semisimple Lie algebra over $\mathcal{R}$ and $g^{¢}$ its complexification. Let $\mathrm{B}(X,Y) = \mathrm{tr}(\mathrm{ad}X\ \mathrm{ad}Y)$ ; $X,Y \in g^{¢}$ be the Killing form. A real form $g_0$ of $g^{¢}$ is called compact if $\mathrm{B}(X,X) < 0$ for each $g_0$ and an automorphism $\theta$ of $g^{¢}$ exists such that

$$\theta g_0 \subset g_0 \quad , \qquad \theta g \subset g$$

and

$$g = k + p \ \ , \ \ g_0 = k + ip$$

where $i = \sqrt{-1}$ , $k$ is the set of all $X \in g$ such that $\theta X = X$ and $p$ is the set of all $Y \in g$ such that $\theta Y = -Y$.

Let $k^{¢}$ and $p^{¢}$ be the subspaces of $g^{¢}$ spanned by $k$, $p$ respectively over ¢. It holds

$$[k^{¢},k^{¢}] \subset k^{¢} \ , \ [k^{¢},p^{¢}] \subset p^{¢} \ , \ [p^{¢},p^{¢}] \subset k^{¢}$$

Let $h$ be a Cartan subalgebra of $g$ and $h^{¢}$ the complexification of $h$. Then $h^{¢}$ is a Cartan subalgebra of $g^{¢}$ and, for the cases considered here (hermitian symmetric spaces of non compact type), holds

$$[h^{¢},k^{¢}] \subset k^{¢} \ , \ [h^{¢},p^{¢}] \subset p^{¢}$$

For given $g^{¢}$, $h^{¢}$, let $\Delta$ be the root system of $g^{¢}$ and $\Delta^{+}$ the system of positive roots. We say that $\alpha$ is compact if $E_\alpha \in k^{¢}$ and non compact if $E_\alpha \in p^{¢}$. The set of compact and non compact roots of $g^{¢}$ with respect to $h^{¢}$ are denoted by $\Delta_c$ and $\Delta_n$ respectively. The set of compact simple roots is denoted by $\Sigma_c$, $\beta$ is the only non compact simple root and $\gamma_r$ is the highest root ( which is a non compact positive root)

Let $k_1 = [k,k]$ and assume that $k$ has a non empty center $\eta$ of dimension one. Then $k = k_1 \oplus \eta$ and $h = (h \leftrightarrow k_1) \oplus \eta$. On the other hand $h^{¢} = (h \leftrightarrow k_1)^{¢} \oplus \eta^{¢}$ is an orthogonal direct sum with respect to the Killing form: for if $H_\mu \in (h \leftrightarrow k_1)^{¢}$ and $H_0 \in h^{¢}$

$$(H_\mu, H_0) = \big([E_\mu, E_{-\mu}], H_0\big) = \big(E_\mu, [E_{-\mu}, H_0]\big) = 0$$

For $\gamma_1,\gamma_2 \in \Delta$ we use the notation

$$\langle \gamma_1,\gamma_2 \rangle = \frac{2(\gamma_1,\gamma_2)}{(\gamma_2,\gamma_2)} = \gamma_1\left(H_{\gamma_2}\right)$$

where (. , .) is the bilinear form on $(h^{¢})^*$ induced by the Killing form on $g^{¢}$.

Let $u(g^{¢})$ be the universal enveloping algebra of $g^{¢}$, $\Lambda \in (h^{¢})^*$ and $R = \frac{1}{2} \sum_{\alpha \in \Delta^+} \alpha$ .

The Verma module $M_\Lambda$ of highest weight $\Lambda$ is defined to be $M_\Lambda = u(g^{¢}) ⊑ I_\Lambda$ where $I_\Lambda$ is the left ideal generated by the elements $(H - \Lambda(H))$ , $H \in h^{¢}$ and the set of generators $X\gamma$ with $\gamma \in \Delta^+$. To fix a basis on $(h^{¢})^*$ we choose the set of compact simple roots $\Sigma_c$ for the space $((h \leftrightarrow k_1)^{¢})^*$ and one element $\varepsilon \in (\eta^{¢})^*$ for which

$$•\varepsilon, \mu ® = 0 \ , \ \forall \mu \in \Sigma_c \text{ and } •\varepsilon, \gamma_r ® = 1 \ ,$$

then each $\Lambda \in (h^{¢})^*$ may be written as $\Lambda = \Lambda_0 + \lambda\varepsilon$ , where $\Lambda_0$ satisfies $•\Lambda, \mu_i ® = •\Lambda_0, \mu_i ®$ $\forall \mu_i \in \Sigma_c$ . If we choose a normalization for $\Lambda_0$ of the type $•\Lambda_0, \gamma_r ® = 0$ , from the last decomposition of $\Lambda$ we conclude that $•\Lambda, \gamma_r ® = \lambda$. The relations $•\Lambda, \mu_i ® = •\Lambda_0, \mu_i ®$ and $•\Lambda_{0,}\gamma_r ® = 0$ fix $\Lambda_0$ uniquely. In the following we consider $\Lambda_0$ to be $k_1$-dominant and integral, that is $•\Lambda_0, \mu_i ® = n_i$ , where $n_i$ are non negative integers.

Now, if $M_\Lambda$ is a Verma module, $\widetilde{L_\Lambda}$ an invariant submodule, and $L_\Lambda = M_\Lambda / \widetilde{L_\Lambda}$ a quotient module and if $\rho_\Lambda = M_\Lambda$ , $\widetilde{L_\Lambda}$ , $L_\Lambda$ is irreducible then we say that $\rho_\Lambda$ is infinitesimally unitary if there exists a scalar product ( , ) on the carrier space $V$ of $\rho_\Lambda$ such that

$$(u, \rho_\Lambda (X) w) = - (\rho_\Lambda (X) u, w)$$

for all $X \in g$ and $u, w \in V$. The above condition is called $g$- invariance.

In a Verma module this scalar product is induced by a sesquilinear form. For definition and construction of this form see Ref [2].

In the following we are going to reformulate the Jakobsen method to calculate the modules $M_\Lambda$ which are unitarizable by using a diagramatic representation of $\Delta_n^+$

## III.- JAKOBSEN METHOD

The modules $M_\Lambda$ are determinated by $\Lambda_0$ and $\lambda$ where $\Lambda_0$ is $k_1$-dominant and integral and $\lambda \in R$

There exists a way to represent the set $\Delta_n^+$ by means of bidimensional diagrams in the following way: one begins with $\beta$ and draws an arrow originating at $\beta$ for each compact simple root $\mu_i$ such that $\beta + \mu_i \in \Delta_n^+$.

Lema 4.1 of Ref [3] shows that $i \leq 2$ . We suppose for simplicity that $i = 2$. Then one draws two arrows: one originating at $\beta + \mu_1$ and parallel to $\mu_2$ and another originating at $\beta + \mu_2$ and parallel to $\mu_1$, both arrows point towards $\beta + \mu_1 + \mu_2$ which is also a root. The next step would be to add compact simple roots to the non compact roots previously obtained by keeping those which are non compact roots. Continuing along these lines the diagram may be completed.

For the description of the possible places for unitarity, Jakobsen uses the Bernshtein, Gel'fand and Gel'fand theorem. This theorem describes the circumstances under which the irreducible quotient $L_\xi$ of a highest weight module can occur in the Jordan-Hölder series JH($M_\Lambda$) of another:

**Definition** : Let $\xi, \Lambda \in (h^{¢})^*$ . A sequence of roots $\alpha_1,..., \alpha_k \in \Delta^+$ is said to satisfy condition (A) for the pair ( $\xi+R$, $\Lambda+R$) if

a) $\xi + R = \sigma_{\alpha_K} ... \sigma_{\alpha\, 1} (\Lambda + R)$ where $\sigma_{\alpha_i}$ is the Weyl reflexion with respect to $\alpha_i$

b) Take $\xi_0 \equiv \Lambda$ , $\xi_i + R = \sigma_{\alpha_i} .. \sigma_{\alpha_i} (\Lambda + R)$

Then $\xi_{i\text{-}1} - \xi_i = n_i\, \alpha_i$ , $n_i \in N$

**Theorem** : (Bernshtein, Gel'fand and Gel'fand); Let $\xi, \Lambda \in (h^{¢})^*$ and let $L_\xi, M_\Lambda$ two Verma modules. Then $L_\xi \in$JH($M_\Lambda$) if and only if there exists a sequence $\alpha_1,...,\alpha_k \in \Delta^+$ satisfying condition (A) for the pair ( $\xi+R$, $\Lambda+R$)

On the other hand, under some conditions the $\alpha_i$ 's may be considered as non compact ones:

**Proposition**: Let $\xi$, $\Lambda \in (h^{\mathbb{C}})^*$ and assume that the sequence $\alpha_1,\ldots, \alpha_k$ satisfies condition (A) for the pair ( $\xi+R$, $\Lambda+R$). If $\xi$ is $k_1$-dominant we may assume that $\alpha_i \in \Delta_n^+$ , $i = 1\ldots k$.

Let $V_{\Lambda_0}$ be an irreducible finite-dimensional $\mathfrak{u}(k_1^{\mathbb{C}})$-module with highest weight $\Lambda_0$ . We first consider the $\mathfrak{u}(k_1^{\mathbb{C}})$-module $p^- \otimes V_{\Lambda_0}$. The highest weights on $p^- \otimes V_{\Lambda_0}$ are of the form $\Lambda_0 - \alpha$ for certain $\alpha \in \Delta_n^+$ which we will describe in terms of the Jakobsen diagrams.

We now describe the method:

**i**) Let $\alpha \in \Delta_n^+$ and assume $\alpha - \mu_j \in \Delta_n^+$ for $\mu_j \in \Sigma_c$ ; $j = 1 ,\ldots, i$ and $i \leq 2$.

Then $\Lambda_0 - \alpha$ is a highest weight for the $\mathfrak{u}(k_1^{\mathbb{C}})$-module $p^- \otimes V_{\Lambda_0}$ if and only if for all $j = 1,\ldots, i$ $\quad \Lambda_0(H_{\mu_j}) \equiv \langle \Lambda_0, \mu_j \rangle \geq \max \{1, \langle \alpha, \mu_j \rangle\}$

Recall that $\Lambda_0$ is fixed by given integers $\langle \Lambda_0, \mu_i \rangle$, $\mu_i \in \Sigma_c$ and $\langle \Lambda_0, \gamma_r \rangle = 0$

**ii**) For those $\alpha \in \Delta_n^+$ of step i) let $\lambda_\alpha \in R$ be determined by the equation

$$\langle \Lambda + R, \alpha \rangle = (\Lambda_0 + \lambda_\alpha \varepsilon + R)(H_\alpha) = 1$$

Let $\lambda_0$ denote the samallest among those $\lambda_\alpha$'s , and let $\alpha_0$ denote the corresponding element of $\Delta_n^+$. We now define the following sets:

$$C^+_{\alpha_0} = \{ \alpha \in \Delta_n^+ / \alpha \geq \alpha_0 \} \quad \text{and} \quad C^-_{\alpha_0} = \{ \alpha \in \Delta_n^+ / \alpha \leq \alpha_0 \}.$$

The way in which those sets appear in the diagram of $\Delta_n^+$ suggest that we can call $C^+_{\alpha_0}$ and $C^-_{\alpha_0}$ the forward and backward cone, respectively, at $\alpha_0$.

**iii**) Let $\omega_q = n_1 \alpha_1 + \ldots + n_r \alpha_r$ , $n_i \in N$. If $\alpha_1, \ldots , \alpha_r \in \Delta_n^+$ satisfies condition (A) for the pair $(\Lambda - \omega_q + R , \Lambda + R)$ where $\Lambda = \Lambda_0 + \lambda_q \varepsilon$ and $\Lambda_0 - \omega_q$ is the weight of a highest weight vector $q$ in the $\mathfrak{u}(k_1^{\mathbb{C}})$-module $\mathfrak{u}(p^-) \otimes V_{\Lambda_0}$ and $\lambda_q < \lambda_0$ , then

$$\alpha_i \in C^+_{\alpha_0} \; \forall i = 1\ldots r$$

**iv**) The $\alpha_i$'s appearing in $\omega_q$ must satisfy certain conditions which we now describe.

Because inner products between positive non compact roots are non negative and because $\lambda_q < \lambda_0$, it follows that for $\alpha_i \in \Delta_n^+$

$$\bullet \Lambda_0 + \lambda_0 \varepsilon + R\ ,\ \alpha_i \circledR > \bullet \Lambda_0 + \lambda_q \varepsilon + R\ ,\ \alpha_i \circledR > 0$$

On the other hand, to check the $k_1$-dominance of $\Lambda_0 - \omega_q$ i.e.

$$\bullet \Lambda_0 - \omega_q\ ,\ \mu_j \circledR \geq 0 \qquad\qquad \forall\ \mu_j \in \Sigma_c$$

is useful to have in mind that if a compact simple root $\mu$ is pointing towards a non compact positive root $\alpha$ in the diagram then $\bullet \alpha , \mu \circledR > 0$ and if $\mu$ arises outwards $\alpha$ then $\bullet \alpha , \mu \circledR \leq 0$

**v)** $M_\Lambda$ with $\Lambda = \Lambda_0 + \lambda_0 \varepsilon$ is unitarizable. The value $\lambda = \lambda_0$ is called the last possible place for unitarity because for $\lambda > \lambda_0$ there is no unitarity. The description of the general situation follows by forming tensor products of $M_\Lambda$ with the unitary module $M_{\lambda_s \varepsilon}$ corresponding to $\Lambda_0 = 0$. The restriction of $M_\Lambda \otimes M_{\lambda_s \varepsilon}$ to the diagonal is the unitarizable module $M_{\Lambda'}$ with $\Lambda' = \Lambda_0 + (\lambda_0 + \lambda_s)\,\varepsilon$ .

This means that if we want to unitarize we must take the quotient space with respect to the invariant subspace generated by the highest weight vector corresponding to $\lambda_0 + \lambda_s$ which is a second order polynomial: this polynomial will be missing.

The modules $M_{\Lambda''}$ with $\Lambda'' = \Lambda_0 + \lambda\,\varepsilon$ , $\lambda_0 + \lambda_s < \lambda < \lambda_0$ are not unitarizable.

For $\lambda_0 + 2\lambda_s$ we may have unitarity and there will be a third order missing polynomial, while there is no unitarity for $\lambda_0 + 2\lambda_s < \lambda < \lambda_0 + \lambda_s$.

Continuing along these lines we arrive at the first possible place for non unitarity which corresponds to $\lambda = \lambda_0 + u\lambda_s$ (we call $u + 1$ the reduction level) and all representations with $\lambda < \lambda_0 + u\lambda_s$ .

The following diagram illustrates the possible places for unitarity (in next section we will see that $\lambda_s < 0$)

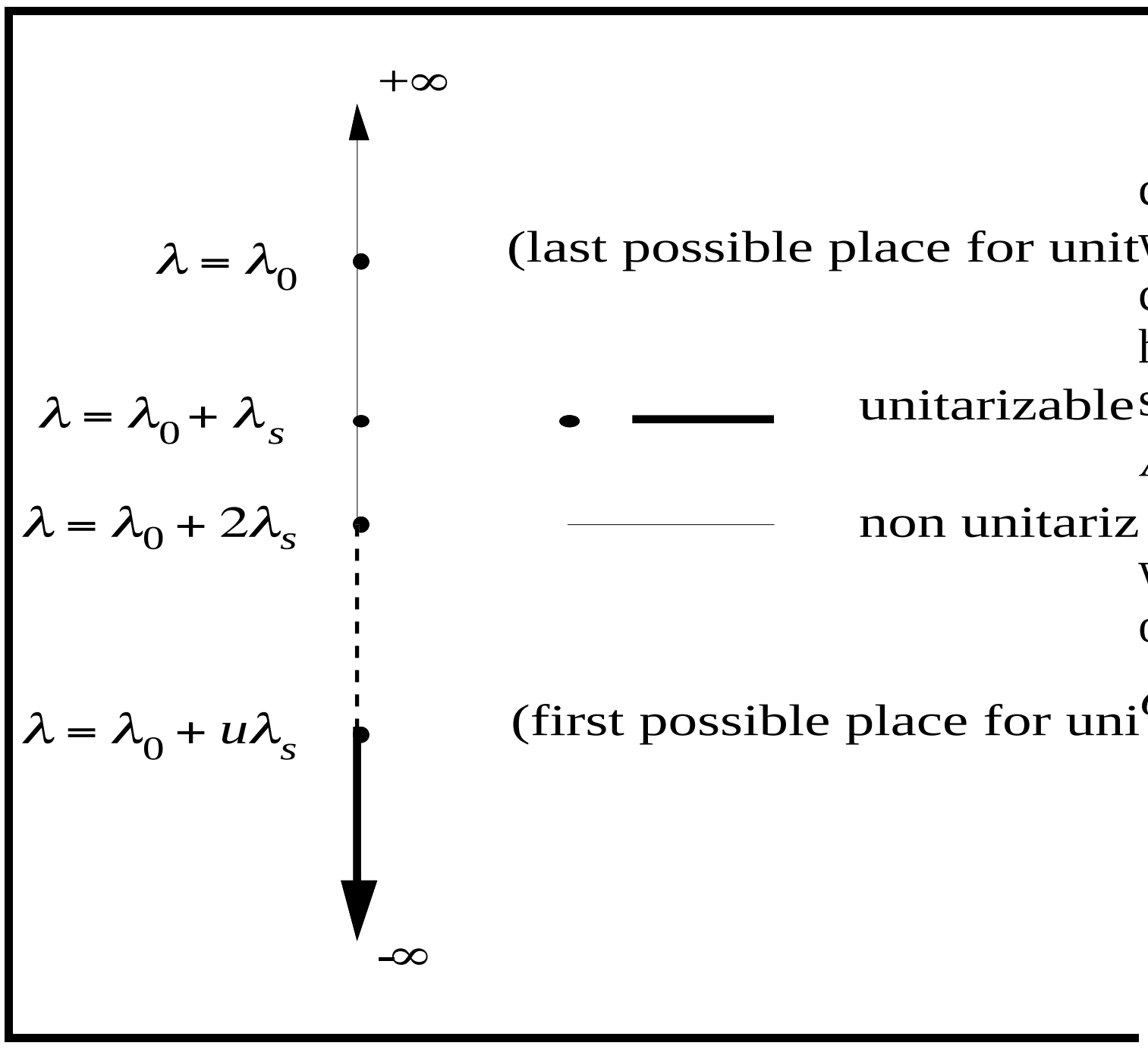

In order to construct the highest weight vector corresponding to the highest weight known, say $\Lambda + R - m_1 \beta - \sum_i m_i \mu_i$ , we start by finding a set of simple roots $\alpha_1, \alpha_2, ...,\alpha_s$ such that

$$\Lambda + R - m_1 \beta - \sum_i m_i \mu_i = \sigma_{\alpha_1} \sigma_{\alpha_2} ... \sigma_{\alpha_s} (\Lambda + R)$$

then we make use of the method outlined in Ref [1]. As can be seen in the examples the exponents are not allways positives, and even in some cases (see $\mathrm{sp}(n,\mathbb{R})$) they are not integers.

The expressions obtained are only formally valid and use has to be made of Taylor series for this powers of the generators and then to apply the commutation relations $\left[\mathrm{E}_{\alpha_i}, \mathrm{E}^{m}_{\alpha_j}\right]$ $(m \in \mathbb{N})$ to each term in the series. For instance, in $\mathrm{su}(2,2)$ from

$\left[\mathrm{E}_{\beta}, \mathrm{E}^{m}_{\mu_1}\right] = \mathrm{m}\, \mathrm{E}^{m-1}_{\mu_1} \mathrm{E}_{\alpha_1}$ ; (where $\beta$ = (0,1,-1,0), $\mu_1$ = (0,0,1,-1) and $\alpha_1$ = (0,1,0,-1))

it follows $\left[\mathrm{E}_{\beta}, \mathrm{f}(\mathrm{E}_{\mu_1})\right] = \mathrm{f}'(\mathrm{E}_{\mu_1})\, \mathrm{E}_{\alpha_1}$ for any analytic function of the operator $\mathrm{E}_{\mu_1}$

## IV.- SOME DEFINITIONS AND NOTATIONS

From the Jakobsen diagrams (see the examples in Section V) we observe that all positive non compact root can be expresed as

$$\alpha = \beta + \mu_{i_1} + ... + \mu_{i_k} \quad , \quad \mu_{i_m} \in \Sigma_c \quad m = 1,...k$$

Thus, given $\alpha$ on this way we define its "height" as $k$+1 (the roots $\mu_{i_m}$ may be repeated)

On the other hand given the decomposition $\Lambda = \Lambda_0 + \lambda\varepsilon$ we may relate the products $\langle \Lambda, \alpha \rangle$ and $\langle \Lambda_0, \alpha \rangle$ , $\alpha \in \Delta_n^+$ , in the following way

$$\langle \Lambda, \alpha \rangle = \langle \Lambda_0, \alpha \rangle + \lambda \frac{(\gamma_r, \gamma_r)}{(\alpha, \alpha)}$$

In fact

$$\langle \Lambda, \alpha \rangle = \langle \Lambda_0, \alpha \rangle + \lambda \langle \varepsilon, \alpha \rangle$$

and, decomposing $\alpha = \gamma_r - \sum_{\mu_\iota \in \Sigma_c} \mu_i$ (see Jakobsen diagrams) then

$$\langle \varepsilon, \alpha \rangle = \frac{2(\varepsilon, \gamma_r)}{(\alpha, \alpha)} = \frac{(\gamma_r, \gamma_r)}{(\alpha, \alpha)}$$

where we use the fact that $\langle \varepsilon, \gamma_r \rangle = 1$ and $\langle \varepsilon, \mu \rangle = 0 \quad \forall \mu \in \Sigma_c$

By means of a direct calculation we obtain the following useful expresions for the products $\langle R, \alpha \rangle$ and $\langle \Lambda, \alpha \rangle$ that will be needed in next section

a) su (p,q) and so*(2n)

$$\langle R, \alpha \rangle = \text{height of } \alpha$$

$$\langle \Lambda, \alpha \rangle = \langle \Lambda_0, \alpha \rangle + \lambda$$

b) sp (n,$\mathbb{R}$)

If $\alpha$ is short

$$\langle R, \alpha \rangle = \text{height} + 1$$

$$\langle \Lambda, \alpha \rangle = \langle \Lambda_0, \alpha \rangle + 2\lambda$$

If $\alpha$ is long

$$\langle R, \alpha \rangle = \frac{1}{2} \{\text{height} + 1\}$$

$$\langle \Lambda, \alpha \rangle = \frac{1}{2} \langle \Lambda_0, \alpha \rangle + \lambda$$

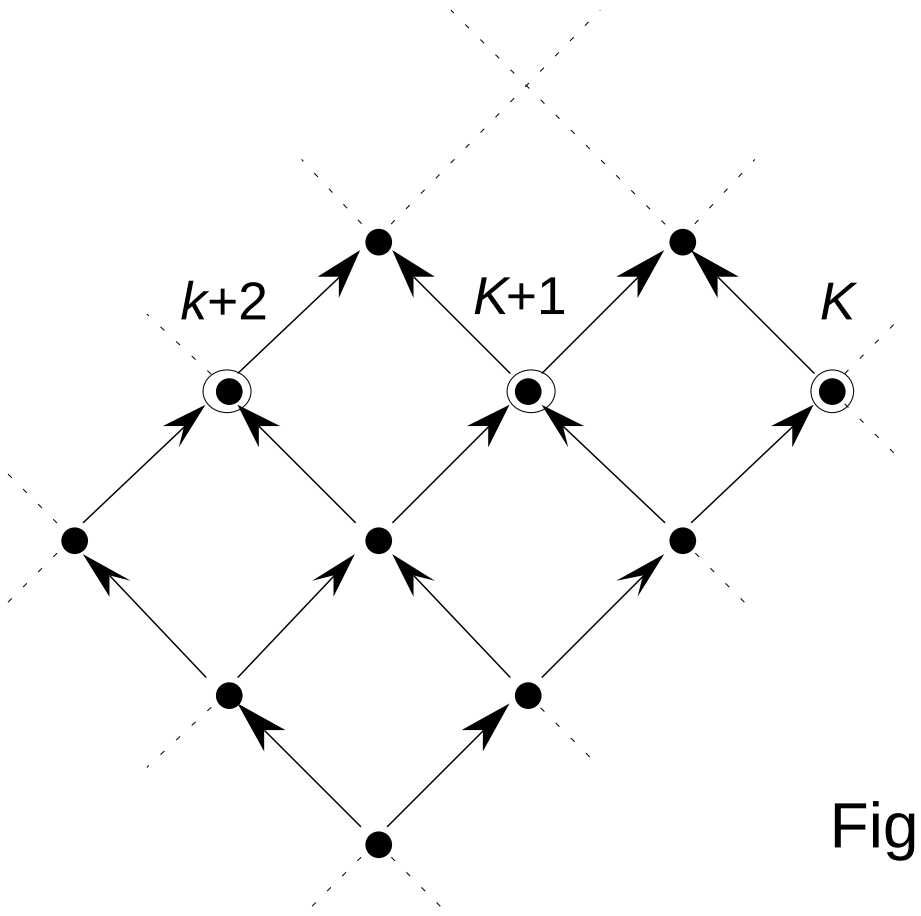

Fig. 1

From the Jakobsen diagrams we see that all roots of the same height are in an horizontal line.
In the figure 1 such a roots are inside a little circle and we may localize them by means of a subindex $j$ which is equal to the height of the root and an ordenation superindex $i$ which is equal to one, for the root placed at the right branch of the cone generated by $\alpha_0$, and it increases from unit to unit when we are going toward the left branch. In this way we will write $\alpha_j^i$ .

In order to calculate the parameter $\lambda_s$ in step **v**) we make use of the following

**Definition** (Harish-Chandra) : Let $\gamma_1$ be the smallest element of $\Delta_n^+$ and, inductively, let $\gamma_k$ be the smallest element of $\Delta_n^+$ which is orthogonal to $\gamma_1, \dots ,\gamma_{k-1}$ . Let $\gamma_1, \dots ,\gamma_t$ the maximal collection obtained. Then $t$ is the split-rank of $g$.

With our notation $\gamma_1 \equiv \beta$. We use the Jakobsen diagrams to obtain the split rank.

## su(p,q)

The collection $\gamma_1, \dots ,\gamma_t$ follows by drawing a line from $\beta$ as is indicated in figure 2.

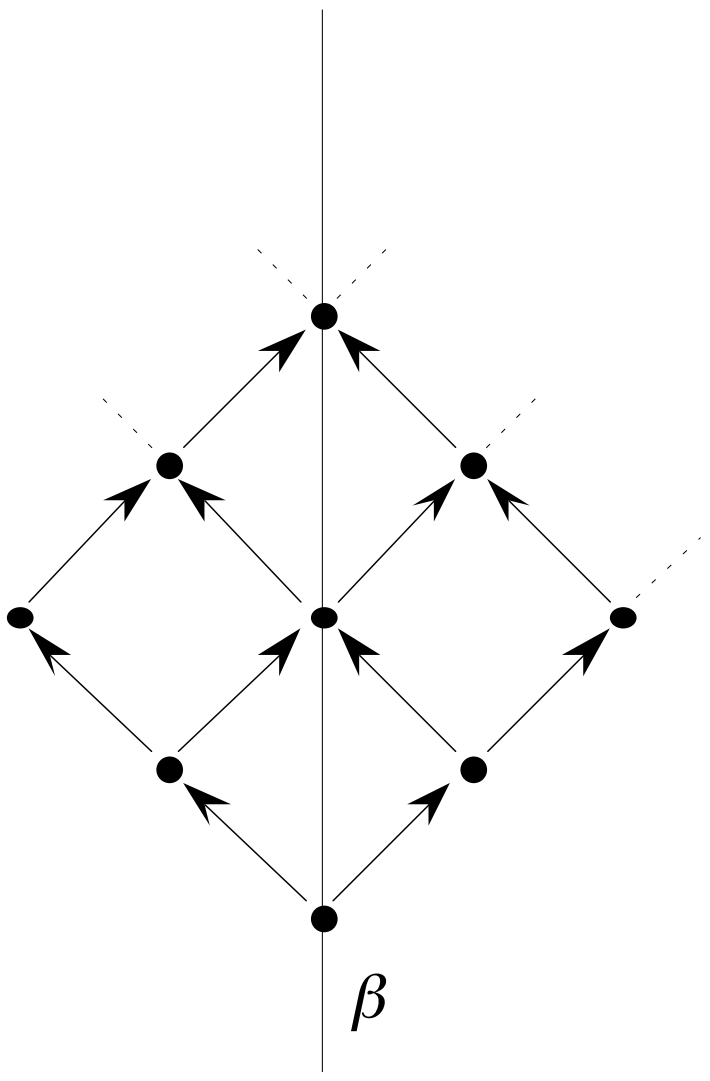
Fig. 2

The roots founded are those which are on the line:

If $p \le q$ : $e_p - e_{p+1}$ , $e_{p-1} - e_{p+2}$ ,...., $e_1 - e_{2p}$
If $p \ge q$ : $e_p - e_{p+1}$ , $e_{p-1} - e_{p+2}$ ,...., $e_{p-(q-1)} - e_{p+q}$

So, if $p \le q$ the split rank is $p$ and if $p \ge q$ the split rank is $q$ , therefore

Split rank su(p, q) = min {p, q}

**sp(n,$R$)**

In the same way as in su(p, q) the collection obtained here is

$$2e_n\ ,\ 2e_{n-1}\ ,...,\ 2e_1$$

Thus

$$\text{Split rank } sp(n,R) = n$$

**so*(2n)**

The collection is, in this case

$$e_{n-1} + e_n\ ,...,e_1 + e_2, \quad \text{if } n \quad \text{is even}$$
$$e_{n-1} + e_n\ ,...,e_2 + e_3, \quad \text{if } n \quad \text{is odd}$$

Then the split rank is $\frac{n}{2}$ if $n$ is even and $\frac{n-1}{2}$ if $n$ is odd:

$$\text{Split rank } so^*(2n) = \left[\frac{n}{2}\right]$$

where $[x]$ denotes the largest integer $\leq x$

**so(2n-1, 2)** , **so(2n-2, 2)**

The split rank is, in both cases, equal to one, because there is no positive non compact root orthogonal to $\beta$

$e_6$ , $e_7$

The collection obtained is now

$$e_6\ :\left\{\ \frac{1}{2}(e_1 - e_2 - e_3 - e_4 - e_5 - e_6 - e_7 + e_8\ ),\ \frac{1}{2}(-e_1 + e_2 + e_3 + e_4 - e_5 - e_6 - e_7 + e_8\ )\right\}$$

then the split rank of $e_6$ is equal to two

$$e_7\ :\left\{e_6 - e_5\ ,\ e_6 + e_5\ ,\ e_8 - e_7\right\}$$

thus split rank of $e_7$ is equal to three.

Now let $\bar{h} = \sum_{i=1}^{t} ¢H_{\gamma}$ and, for $1 \leq j \leq t$ , $t$ being the split rank, let $c_j$ be the number of compact positive roots $\mu$ such that $\mu^{™}{}_{\bar{h}} = \frac{1}{2}(\gamma_j - \gamma_i)$ , $i < j$ . Then, if we consider the most singular non trivial unitary module corresponding to $\Lambda_0 = 0$, according to Theorem 5.10 in Ref [4] $\lambda_q = -\frac{1}{2} c_j$ . A straightforward calculation case by case shows that

$$\lambda_q = (j - 1)\, \lambda_s \quad , \quad 1 \leq j \leq t$$

with $\lambda_s$ given in the following table

| | $su(p,q)$ | $sp(n, \Re)$ | $so^*(2n)$ | $e_6$ | $e_7$ |
|---|---|---|---|---|---|
| $\lambda_s$ | -1 | $-\frac{1}{2}$ | -2 | -3 | -4 |

## V.- POSSIBLE PLACES FOR UNITARITY

We consider here those cases for which the reduction level is strictly higher than one.

### su(p,q)

Dynkin diagram

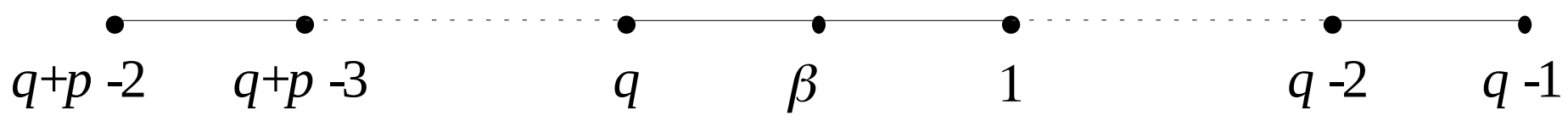

Let $M_\Lambda$ be a representation for su(p, q) with $\Lambda = (\Lambda_1, \Lambda_2 , ..., \Lambda_{p+q} )$ with respect to the standard orthonormal basis of $R^n$ $(n = p + q)$ , satisfying the following conditions on its components:

$$\Lambda_1 = \Lambda_2 = ... = \Lambda_i \neq \Lambda_{i+1}$$
$$\Lambda_n = \Lambda_{n-1} = ... = \Lambda_{n-j+1} \neq \Lambda_{n-j}$$

If we put $\Lambda = \Lambda_0 + \varepsilon\lambda$ these conditions are equivalent to the following ones:

$$\bullet\Lambda_0 , \mu_{q-1}\circledR = \bullet\Lambda_0 , \mu_{q-2}\circledR =....= \bullet\Lambda_0 , \mu_{t+1}\circledR = 0, \quad \bullet\Lambda_0 , \mu_t \circledR \neq 0$$
$$\bullet\Lambda_0 , \mu_{n-2}\circledR = \bullet\Lambda_0 , \mu_{n-3}\circledR = ... = \bullet\Lambda_0 , \mu_{s+1}\circledR = 0, \quad \bullet\Lambda_0 , \mu_s \circledR \neq 0$$

with $t = q - j$ and $s = n - i - 1$

Applying steps **i**) and **ii**) we obtain

$$\alpha_0 = \beta + \mu_1 + ... + \mu_t + \mu_q + \mu_{q+1} + ... + \mu_s$$

with height $t + s - q + 2$ . Then $\lambda_0 = q - t - s - 1$

Then a first order polynomial will be missing with highest weight

$$\Lambda_0 + ( q - t - s - 1)\varepsilon + R - \alpha_0$$

where, in this case, $\varepsilon = (q/n , q/n , ... ,q/n , -p/n , ... ,-p/n )$ with $p$ copies of $q/n$ and $q$ of $-p/n$ .

For $\lambda_q = \lambda_0 + \lambda_s = \lambda_0 - 1$ we obtain from steps **iii**) and **iv**) a second order polynomial that will be missing with heighest weight

$$\Lambda_0 + (\lambda_0 - 2)\varepsilon + R - \overset{1}{\alpha}_{2-\lambda_0} - \overset{2}{\alpha}_{2-\lambda_0}.$$

Next case is $\lambda_q = \lambda_0 - 2$ where a third order polynomial with highest weight $\Lambda_0 + (\lambda_0 - 2)\varepsilon + R - \overset{1}{\alpha}_{3-\alpha_0} - \overset{2}{\alpha}_{3-\lambda_0} - \overset{3}{\alpha}_{3-\lambda_0}$ will be missing. Continuing in the same way we arrive at $\lambda_q = \lambda_0 - u$, u = min ($q - t - 1$, $n - s - 2$) where a polynomial of order $u + 1$ = min ($i$, $j$) will be missing. For $\lambda < \lambda_0 - u$ it is impossible to find a polynomial of order strictly higher than min ($i$, $j$) because the $k_1$-dominance is violated (see step **iv**).

Thus the reduction level is min ($i$, $j$ ). On the other hand $\lambda = \bullet\Lambda, \gamma_r$ ® or, equivalently, $\Lambda_n = \Lambda_1 - \lambda$ . Then, taking into account the possible values of $\lambda$ we obtain the following diagram:

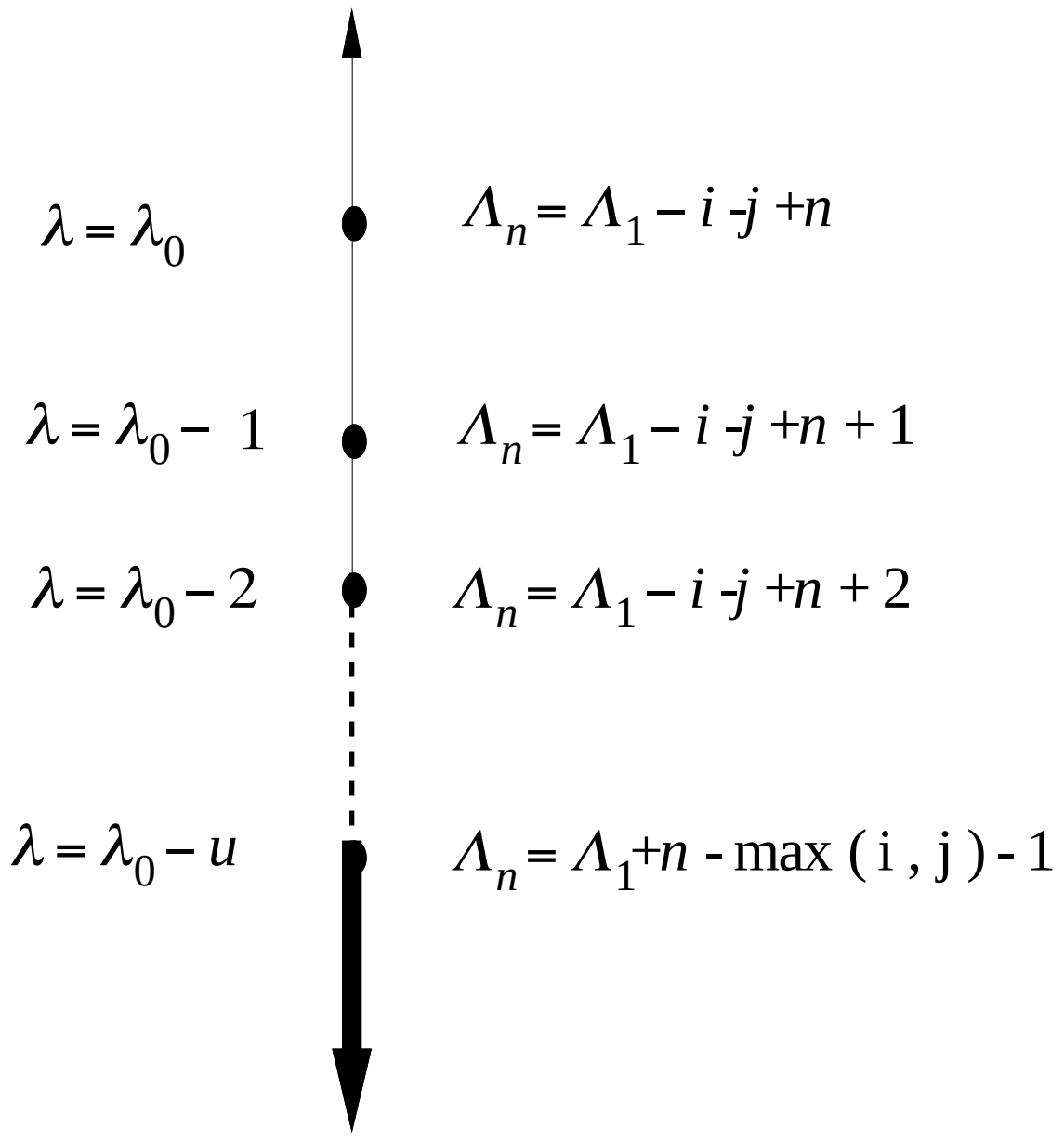

**Example: su(5,8)**

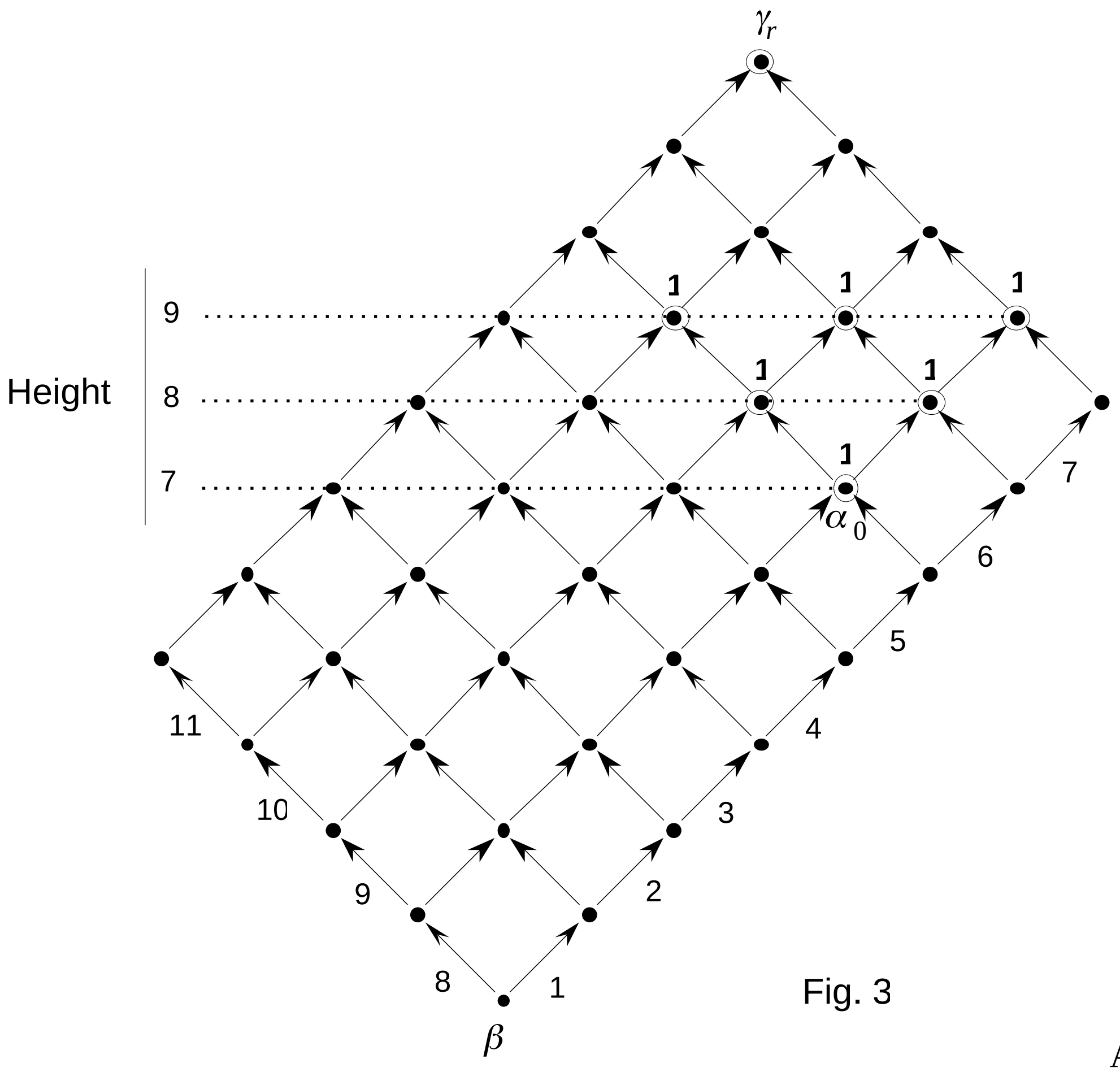

Fig. 3

Assume that

$$\langle \Lambda_0, \mu_7 \rangle = \langle \Lambda_0, \mu_6 \rangle = 0$$

$$\langle \Lambda_0, \mu_5 \rangle \neq 0$$

$$\langle \Lambda_0, \mu_{11} \rangle = \langle \Lambda_0, \mu_{10} \rangle = \langle \Lambda_0, \mu_9 \rangle = 0$$

$$\langle \Lambda_0, \mu_8 \rangle \neq 0$$

$$\langle \Lambda_0, \mu_i \rangle = n_i \quad ,, \quad 1 \leq i \leq 5 \quad \text{or} \quad i = 8$$

With those conditions we obtain $\alpha_0 = \beta + \mu_1 + \mu_2 + \mu_3 + \mu_4 + \mu_5 + \mu_8$

the height of which is 7 , then $1 = \langle \Lambda, \alpha_0 \rangle + \langle R, \alpha_0 \rangle = \langle \Lambda, \alpha_0 \rangle + 7 = \lambda_0 + 7$ or $\lambda_0 = -6$

For $\lambda_q = \lambda_0 + \lambda_s = -7$ and having in mind that now $\Lambda' = \Lambda_0 - 7\varepsilon$ :

$$\bullet \Lambda + R, \alpha^1{}_8 \circledR = -7 + \bullet R, \alpha^1{}_8 \circledR = 1$$

$$\bullet \Lambda + R, \alpha^2{}_8 \circledR = 1$$

then a second order polynomial will be missing with highest weight $\Lambda' + R - \alpha^1{}_8 - \alpha^2{}_8$ .

For $\lambda_q = -8$ ,, $\Lambda'' = \Lambda_0 - 8\varepsilon$ and we have

$$\bullet \Lambda + R, \alpha^1{}_9 \circledR = \bullet \Lambda + R, \alpha^2{}_9 \circledR = \bullet \Lambda + R, \alpha^3{}_9 \circledR = 1$$

then a third order polynomial will be missing with highest weight $\Lambda'' + R - \alpha^1{}_9 - \alpha^2{}_9 - \alpha^3{}_9$ .

For $\lambda_q < -8$ only a set of roots belonging to $C^+_{\alpha_{(}}$ and without a coefficient equal to one in the Fig. 3 could satisfy condition (A). But, for example, $\alpha^1{}_{10}$ could not belong to this set because of $\mu_9$ which point towards it and because there is no root between those from which arises $\mu_9$, the result would not be $k_1$-dominant. The same thing occurs with $\alpha^2{}_{10}$ , $\alpha^1{}_{11}$ because of $\mu_{10}$ and with $\alpha^3{}_{10}$ , $\alpha^2{}_{11}$ , $\gamma_r$ because of $\mu_{11}$ . There is unitarity for $\lambda_q < -8$ .

The highest weight vectors which we must eliminate (missing polynomials) in order to obtain unitarity are, formally, the followings:

**1)** Height 7

$$E_{-\mu_5}^{-n_5} E_{-\mu_4}^{-n_4-n_5-1} E_{-\mu_3}^{-n_3-n_4-n_5-2} E_{-\mu_2}^{-n_2-n_3-n_4-n_5-3}$$

$$\times E_{-\mu_1}^{-n_1-n_2-n_3-n_4-n_5-4} E_{-\mu_8}^{-n_8} E_{-\beta} E_{-\mu_8}^{n_8+1} E_{-\mu_1}^{n_1+n_2+n_3+n_4+n_5+5}$$

$$\times E_{-\mu_2}^{n_2+n_3+n_4+n_5+4} E_{-\mu_3}^{n_3+n_4+n_5+3} E_{-\mu_4}^{n_4+n_5+2} E_{-\mu_5}^{n_5+1}$$

with $\Lambda = \Lambda_0 - 6\varepsilon + R - \beta - \mu_1 - \mu_2 - \mu_3 - \mu_4 - \mu_5 - \mu_8$

**2)** Height 8

$$E_{-\mu_5}^{-n_5}\, E_{-\mu_4}^{-n_4-n_5-1}\, E_{-\mu_3}^{-n_3-n_4-n_5-2}\, E_{-\mu_2}^{-n_2-n_3-n_4-n_5-3}$$

$$\times\, E_{-\mu_1}^{-n_1-n_2-n_3-n_4-n_5-4}\, E_{-\mu_8}^{-n_8}\, E_{-\beta}^{2}\, E_{-\mu_8}^{n_8+2}\, E_{-\mu_9}$$

$$\times\, E_{-\mu_1}^{n_1+n_2+n_3+n_4+n_5+6}\, E_{-\mu_2}^{n_2+n_3+n_4+n_5+5}\, E_{-\mu_3}^{n_3+n_4+n_5+4}$$

$$\times\, E_{-\mu_4}^{n_4+n_5+3}\, E_{-\mu_5}^{n_5+2}\, E_{-\mu_6}$$

with $\Lambda = \Lambda_0 - 7\varepsilon + R - 2\beta - 2\mu_1 - 2\mu_2 - 2\mu_3 - 2\mu_4 - 2\mu_5 - \mu_6 - 2\mu_8 - \mu_9$

**3)** Height 9

$$E_{-\mu_5}^{-n_5}\, E_{-\mu_4}^{-n_5-n_4-1}\, E_{-\mu_3}^{-n_3-n_4-n_5-2}\, E_{-\mu_2}^{-n_2-n_3-n_4-n_5-3}\, E_{-\mu_1}^{-n_1-n_2-n_3-n_4-n_5-4}$$

$$\times\, E_{-\mu_8}^{-n_8}\, E_{-\beta}^{3}\, E_{-\mu_8}^{n_8+3}\, E_{-\mu_9}^{2}\, E_{-\mu_{10}}\, E_{-\mu_1}^{n_1+n_2+n_3+n_4+n_5+7}$$

$$\times\, E_{-\mu_2}^{n_2+n_3+n_4+n_5+6}\, E_{-\mu_3}^{n_3+n_4+n_5+5}\, E_{-\mu_4}^{n_4+n_5+4}\, E_{-\mu_5}^{n_5+3}\, E_{-\mu_6}^{2}\, E_{-\mu_7}$$

with $\Lambda = \Lambda_0 - 8\varepsilon + R - 3\beta - 3\mu_1 - 3\mu_2 - 3\mu_3 - 3\mu_4 - 3\mu_5 - 2\mu_6 - \mu_7 - 3\mu_8 - 2\mu_9 - \mu_{10}$

In the three cases, in order to cancel the negative powers of the generators, appropiate commutation relations are to be applied, as stated before.

## sp $(n, R)$

Dynkin diagram

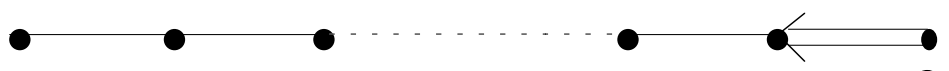

Let $M_\Lambda$ be a representation for sp $(n, R)$ with $\Lambda = (\Lambda_1, \dots, \Lambda_n)$ we put $\Lambda = \Lambda_0 + \lambda\varepsilon$ where $\varepsilon = (1, 1, \dots, 1)$. We consider two cases:

**<u>Case I</u>**

The weight $\Lambda$ satisfy the following conditions on its components

$$\Lambda_1 = \Lambda_2 = \dots = \Lambda_i \geq \Lambda_{i+1} + 2$$

or, equivalently

$$\langle \Lambda_0, \mu_1 \rangle = \langle \Lambda_0, \mu_2 \rangle = \dots = \langle \Lambda_0, \mu_{i-1} \rangle = 0 \ , \ \langle \Lambda_0, \mu_i \rangle = n \geq 2$$

Applying Jakobsen method we obtain

$$\alpha_0 = \beta + 2\mu_{n-1} + 2\mu_{n-2} + \dots + 2\mu_{n-(n-i)}$$

with height $2(n-i)+1$. As $\alpha_0$ is a long root, the condition $\langle \Lambda + R, \alpha_0 \rangle = 1$ implies

$$\frac{1}{2}(\Lambda_0, \alpha_0) + \lambda_0 + n - i + 1 = 1 \quad , \quad \lambda_0 = i - n$$

then a first order polynomial with highest weight $\Lambda_0 + (i-n)\varepsilon + R - \alpha_0$ will be missing when we unitarize.

For $\lambda = \lambda_0 + \lambda_s = i - n - \frac{1}{2}$ we obtain a second order polynomial which will be missing with highest weight

$$\Lambda_0 + (i - n - \tfrac{1}{2})\varepsilon + R - 2\alpha^1_{2(n-i)+2}$$

For $\lambda = \lambda_0 + 2\lambda_s$ the third order missing polynomial has highest weight

$$\Lambda_0 + (i - n - 1)\varepsilon + R - 2\alpha^1_{2(n-i)+3} - \alpha^2_{2(n-i)+3}$$

Following along these lines we arrive at $\lambda = \lambda_0 + (i-1)\lambda_s = \frac{1}{2}(i+1) - n$ where a ith order polynomial will be missing. For $\lambda < \frac{1}{2}(i+1) - n$ it is impossible to obtain polynomials of order strictly higher than $i$ because there would not be $k_1$-dominance, therefore the reduction level is $i$. On the other hand from the condition $\lambda = \langle \Lambda, \gamma_r \rangle$ it follows that $\Lambda_1 = \lambda$ Thus, for the different values of $\lambda$ we obtain the following diagram which give us the possible values of $\Lambda_1$ for unitarity:

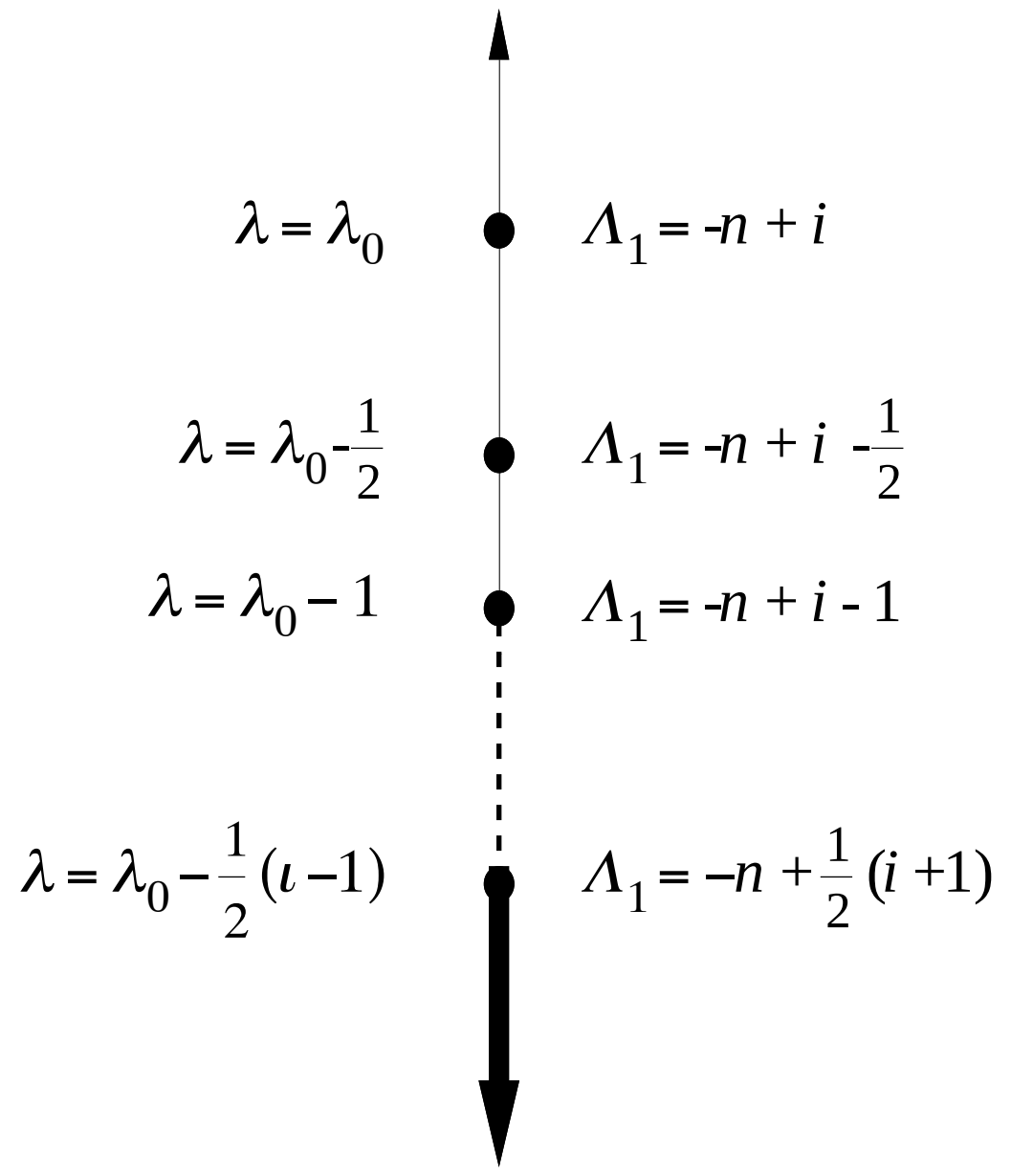

**Case II**

We consider in this case the following conditions

$$\Lambda_1 = \Lambda_2 = \ldots = \Lambda_i \quad ; \quad \Lambda_i - \Lambda_{i+1} = 1 \ , \ \Lambda_{i+1} = \Lambda_{i+2} = \ldots = \Lambda_{i+j} \neq \Lambda_{i+j+1}$$

which are equivalent to the following ones:

$$\langle \Lambda_0 , \mu_1 \rangle = \langle \Lambda_0 , \mu_2 \rangle = \ldots = \langle \Lambda_0 , \mu_{i-1} \rangle = 0 \quad , \quad \langle \Lambda_0 , \mu_i \rangle = 1$$

$$\langle \Lambda_0 , \mu_{i+1} \rangle = \langle \Lambda_0 , \mu_{i+2} \rangle = \ldots = \langle \Lambda_0 , \mu_{i+j-1} \rangle = 0 \quad , \quad \langle \Lambda_0 , \mu_{i+j} \rangle = n \quad \geq 1$$

In this case

$$\alpha_0 = \beta + 2 \left( \mu_{n-1} + \mu_{n-2} + \ldots + \mu_{n-(n-i-j)} \right) + \mu_{n-(n-i-j)} + \ldots + \mu_{n-(n-i)}$$

the height of which is $2(n-i) - j + 1$ . As $\alpha_0$ is a short root, the condition $\langle \Lambda + R , \alpha_0 \rangle = 1$ implies

$$(\Lambda_0, \alpha_0) + 2\lambda_0 + 2(n-i+1) - j = 1$$

and, having in mind that we can also state

$$\alpha_0 = \gamma_r - 2\,(\mu_1 + \ldots + \mu_{i-1}) - \mu_i - \mu_{i+1} - \ldots - \mu_{i+j-1}$$

we have

$$\lambda_0 = i - n + \frac{j}{2}$$

For $\lambda = \lambda_0 + \lambda_s = i - n + \frac{(j-1)}{2}$ we obtain a second order polynomial which will be missing with highest weight

$$\Lambda_0 + \left(i - n + \frac{(j-1)}{2}\right)\varepsilon + R - \alpha^1_{2(n-i+1)-j} - \alpha^2_{2(n-i+1)-j}$$

Continuing as in case I we arrive at $\lambda = \lambda_0 + (i-1)\lambda_s = -n + \frac{1}{2}(i + j + 1)$ where a ith order polynomial will be missing. As for $\lambda < -n + \frac{1}{2}(i + j + 1)$ there is no $k_1$-dominance the reduction level is $i$ . The diagram in this case is the following:

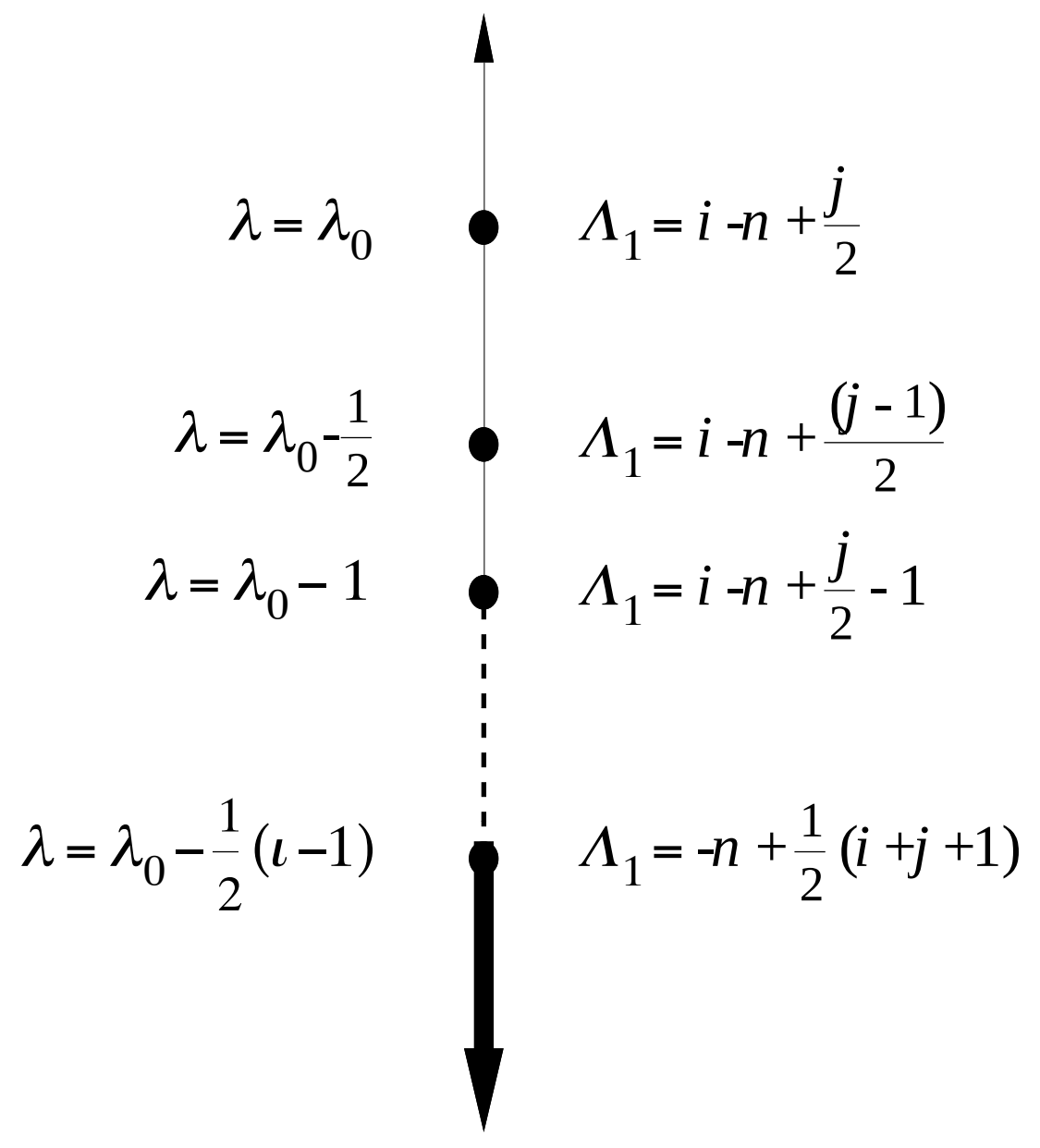

**Example: sp (10,$R$)**

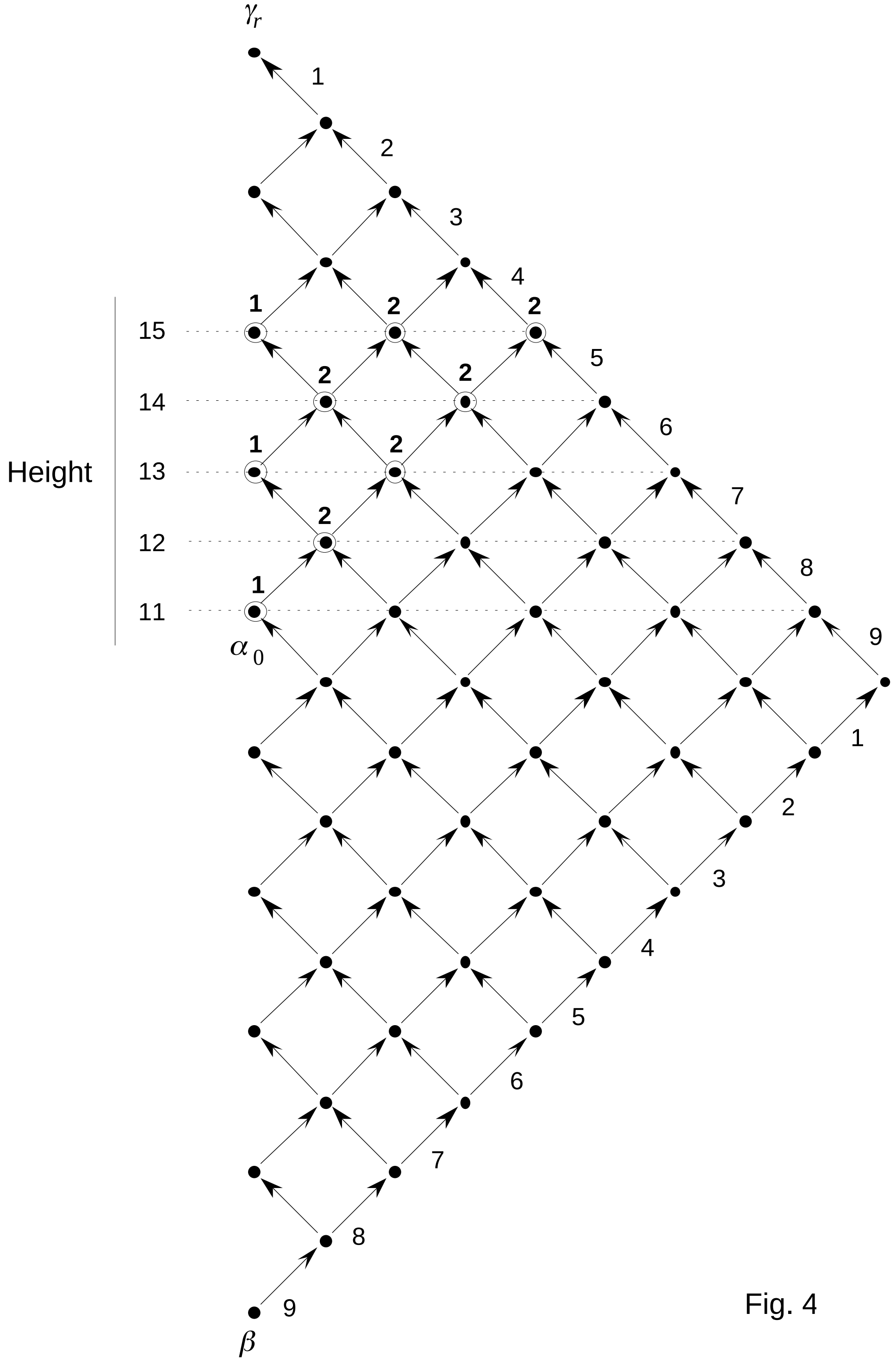

Fig. 4

**<u>Case I</u>**

Let be the following conditions on $\Lambda_0$

$$•\Lambda_0,\ \mu_1® = •\Lambda_0,\ \mu_2® = •\Lambda_0,\ \mu_3® = •\Lambda_0,\ \mu_4® = 0$$

$$\bullet\Lambda_0, \mu_5 \circledR = n_5 \geq 2 \qquad \bullet\Lambda_0, \mu_i \circledR = n_i \qquad 6 \leq i \leq 9$$

In this case

$$\alpha_0 = \beta + 2\mu_9 + 2\mu_8 + 2\mu_7 + 2\mu_6 + 2\mu_5$$

with height 11

As $\alpha_0$ is a long root, the condition $\bullet\Lambda + R, \alpha_0 \circledR = 1$ implies

$$\frac{1}{2}(\Lambda_0, \alpha_0) + \lambda_0 + 6 = 1 \; ; \quad \lambda_0 = -5$$

For $\lambda_q = \lambda_0 + \lambda_s = -\frac{11}{2}$

$$\bullet\Lambda + R\, , \alpha^1_{12} \circledR = 2\lambda_q + 13 = 2 \text{ because } \alpha^1_{12} \text{ is short}$$

Then there is a second order polynomial missing for $\lambda_q = -\frac{11}{2}$ with highest weight $\Lambda_0 - \frac{11}{2}\varepsilon + R - 2\alpha^1_{12}$

For $\lambda_q = \lambda_0 + 2\lambda_s = -6$

$$\bullet\Lambda + R, \alpha^1_{13} \circledR = 2\lambda_q + 14 = 2 \text{ because } \alpha^1_{13} \text{ is short}$$

$$\bullet\Lambda + R, \alpha^2_{13} \circledR = \lambda_q + 7 = 1 \text{ because } \alpha^2_{13} \text{ is long}$$

and we will have a third order missing polynomial with highest weight $\Lambda_0 - 6\varepsilon + R - 2\alpha^1_{13} - \alpha^2_{13}$

For $\lambda_q = -\frac{13}{2}$

$$\bullet\Lambda + R, \alpha^1_{14} \circledR = \bullet\Lambda + R, \alpha^2_{14} \circledR = 2\lambda_q + 15 = 2$$

a fourth order polynomial will be missing with highest weight $\Lambda_0 - \frac{13}{2}\varepsilon + R - 2\alpha^1_{14} - 2\alpha^2_{14}$

For $\lambda_q = -7$

$$\bullet\Lambda + R, \alpha^1_{15} \circledR = \bullet\Lambda + R, \alpha^2_{15} \circledR = 2\lambda_q + 16 = 2$$

$$\bullet\Lambda + R, \alpha^3_{15} \circledR = 2\lambda_q + 8 = 1$$

and there will be a fifth order polynomial missing with highest weight

$$\Lambda_0 - 7\varepsilon + R - 2\alpha^{1}_{15} - 2\alpha^{2}_{15} - \alpha^{3}_{15}$$

For $\lambda_q < -7$ the roots which we must consider are those belonging to $C_{\alpha_0}^{+}$ without a coefficient in the figure 4.

By an argument along the lines of the example for su (5, 8) we see that for those $\lambda_q$ there is no $k_1$ -dominance. Then the first possible place for non unitarity is $\lambda_q = -7$

We state in the following the highest weight vectors for the two first heights.

**1)** Height 11

$$E_{-\mu_5}^{-n_5+1}\; E_{-\mu_6}^{-n_5-n_6}\; E_{-\mu_7}^{-n_5-n_6-n_7-1}\; E_{-\mu_8}^{-n_5-n_6-n_7-n_8-2}$$

$$\times\; E_{-\mu_9}^{-n_5-n_6-n_7-n_8-n_9-3}\; E_{-\beta}\; E_{-\mu_9}^{n_5+n_6+n_7+n_8+n_9+5}$$

$$\times\; E_{-\mu_8}^{n_5+n_6+n_7+n_8+4}\; E_{-\mu_7}^{n_5+n_6+n_7+3}\; E_{-\mu_6}^{n_5+n_6+2}\; E_{-\mu_5}^{n_5+1}$$

with $\Lambda = \Lambda_0 - 5\varepsilon + R - \beta - 2\mu_5 - 2\mu_6 - 2\mu_7 - 2\mu_8 - 2\mu_9$

**2)** Height 12

$$E_{-\mu_5}^{-n_5+1}\; E_{-\mu_6}^{-n_5-n_6}\; E_{-\mu_7}^{-n_5-n_6-n_7-1}\; E_{-\mu_8}^{-n_5-n_6-n_7-n_8-2}\; E_{-\mu_9}^{-n_5-n_6-n_7-n_8-n_9-3}$$

$$\times\; E_{-\beta}^{3/2}\; E_{-\mu_9}^{n_5+n_6+n_7+n_8+n_9+6}\; E_{-\mu_8}^{n_5+n_6+n_7+n_8+5}\; E_{-\mu_7}^{n_5+n_6+n_7+4}\; E_{-\mu_6}^{n_5+n_6+3}\; E_{-\mu_5}^{n_5+2}$$

$$\times\; E_{-\mu_4}^{2}\; E_{-\mu_5}^{-n_5}\; E_{-\mu_6}^{-n_5-n_6-1}\; E_{-\mu_7}^{-n_5-n_6-n_7-2}\; E_{-\mu_8}^{-n_5-n_6-n_7-n_8-3}\; E_{-\mu_9}^{-n_5-n_6-n_7-n_8-n_9-4}$$

$$\times\; E_{-\beta}^{1/2}\; E_{-\mu_9}^{n_5+n_6+n_7+n_8+n_9+5}\; E_{-\mu_8}^{n_5+n_6+n_7+n_8+4}\; E_{-\mu_7}^{n_5+n_6+n_7+3}\; E_{-\mu_6}^{n_5+n_6+2}\; E_{-\mu_5}^{n_5+1}$$

with $\Lambda = \Lambda_0 - \frac{11}{2}\varepsilon + R - 2\beta - 2\mu_4 - 4\mu_5 - 4\mu_6 - 4\mu_7 - 4\mu_8 - 4\mu_9$

**<u>Case II</u>**

Let now be the following conditions in $\Lambda_0$

$$\langle \Lambda_0, \mu_1 \rangle = \langle \Lambda_0, \mu_2 \rangle = 0 \qquad \langle \Lambda_0, \mu_3 \rangle = \langle \Lambda_0, \mu_4 \rangle = 1$$

•$\Lambda_0, \mu_5$® = •$\Lambda_0, \mu_6$® = 0 •$\Lambda_0, \mu_7$® = $n$ ≥ 1

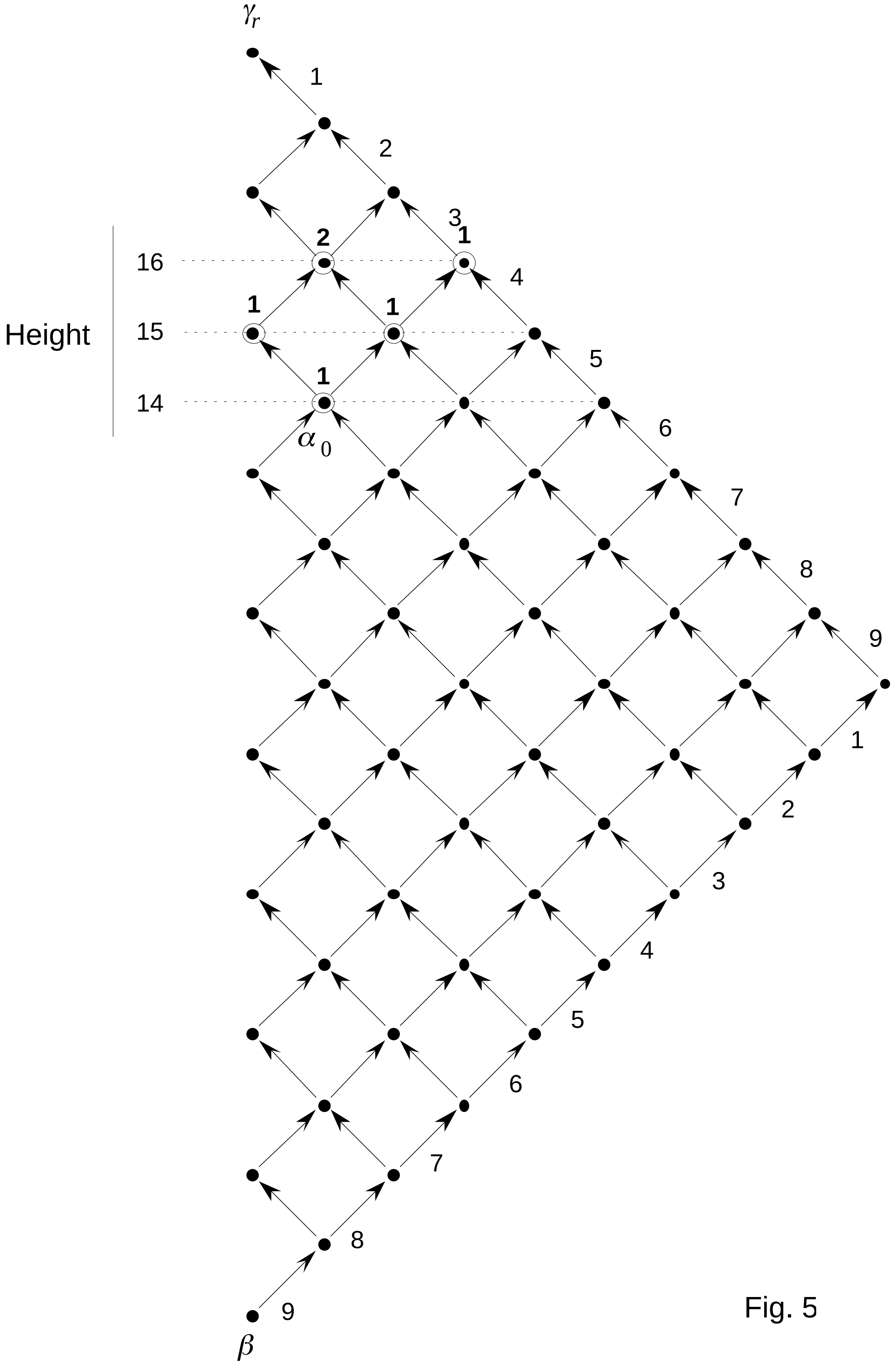

Fig. 5

In this case $\alpha_0 = \beta + 2\mu_9 + 2\mu_8 + 2\mu_7 + 2\mu_6 + 2\mu_5 + 2\mu_4 + \mu_3$ with height 14. As $\alpha_0$ is a short root:

$$1 = •\Lambda + R, \alpha_0 ® = (\Lambda_0, \gamma_r \ -2\mu_1 - 2\mu_2 - \mu_3) + 2\lambda_0 + 15 \quad ; \quad \lambda_0 = -\frac{13}{2}$$

Para $\lambda_q = \lambda_0 + \lambda_s = -7$

$$\bullet \Lambda + R, \alpha^1_{15} ® = 2\lambda_q + 15 = 1$$

$$\bullet \Lambda + R, \alpha^2_{15} ® = \lambda_q + 8 = 1$$

then a second order polynomial with highest weight $\Lambda_0 - 7\varepsilon + R - \alpha^1_{15} - \alpha^2_{15}$ will be missing.

For $\lambda_q = \lambda_0 + 2\lambda_s = -\frac{15}{2}$

$$\bullet \Lambda + R, \alpha^1_{16} ® = 2\lambda_q + 16 = 1$$

$$\bullet \Lambda + R, \alpha^2_{16} ® = 2\lambda_q + 17 = 2$$

then a third order polynomial will be missing with highest weight $\Lambda_0 - \frac{15}{2}\varepsilon + R - \alpha^1_{16} - 2\alpha^2_{16}$

For $\lambda_q < -\frac{15}{2}$ there exists unitarity.

**so* (2n)**

Dynkin diagram

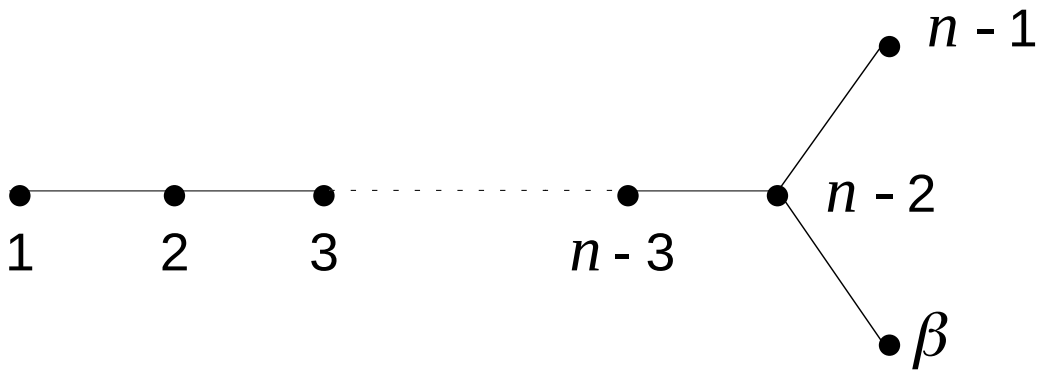

Let now $M_\Lambda$ be a representation for so* (2n) with $\Lambda = (\Lambda_1, \dots, \Lambda_n)$. In this case given $\Lambda = \Lambda_0 + \lambda\varepsilon$ we have $\varepsilon = (\frac{1}{2}, \frac{1}{2}, \dots, \frac{1}{2})$. We consider the following conditions on its components

$$\Lambda_1 = \Lambda_2 = \dots = \Lambda_i > \Lambda_{i+1} + 1 \qquad i \neq 1$$

or, equivalently

•$\Lambda_0, \mu_1$® = •$\Lambda_0, \mu_2$® = ....= •$\Lambda_0, \mu_{i-1}$® = 0 , •$\Lambda_0, \mu_i$® = $n > 1$

From the Jakobsen method we have

$$\alpha_0 = \beta + (\mu_{n-1} + \mu_{n-2}) + (\mu_{n-2} + \mu_{n-3}) + \dots + (\mu_{n-(n-i)} + \mu_{n-(n-i)-1})$$

the height of which is $2(n - i) + 1$. The condition •$\Lambda + R, \alpha_0$® = 1 implies in this case

$$(\Lambda_0, \alpha_0) + \lambda_0 + 2(n - i) + 1 = 1 \quad ; \quad \lambda_0 = 2(i - n)$$

For $\lambda = \lambda_0 + \lambda_s = 2(i - n - 1)$ we obtain a second order polynomial which will be missing with highest weight

$$\Lambda_0 + 2(i - n - 1)\varepsilon + R - \alpha^1_{2(n-i)+3} - \alpha^2_{2(n-i)+3}$$

Following along these lines we arrive at $\lambda = \lambda_0 + \{[\frac{i}{2}] - 1\}\lambda_s$ where a polynomial with order $[\frac{i}{2}]$ is missing. For $\lambda < 2(i - n - \{[\frac{i}{2}] - 1\})$ there is impossible to obtain missing polynomials the order of which is strictly higher than $[\frac{i}{2}]$ because in those places the weights are not $k_1$-dominants then the reduction level is $[\frac{i}{2}]$. From the condition $\lambda$ = •$\Lambda, \gamma_r$® we obtain $\Lambda_1 = \frac{\lambda}{2}$. In this way we obtain the following diagram

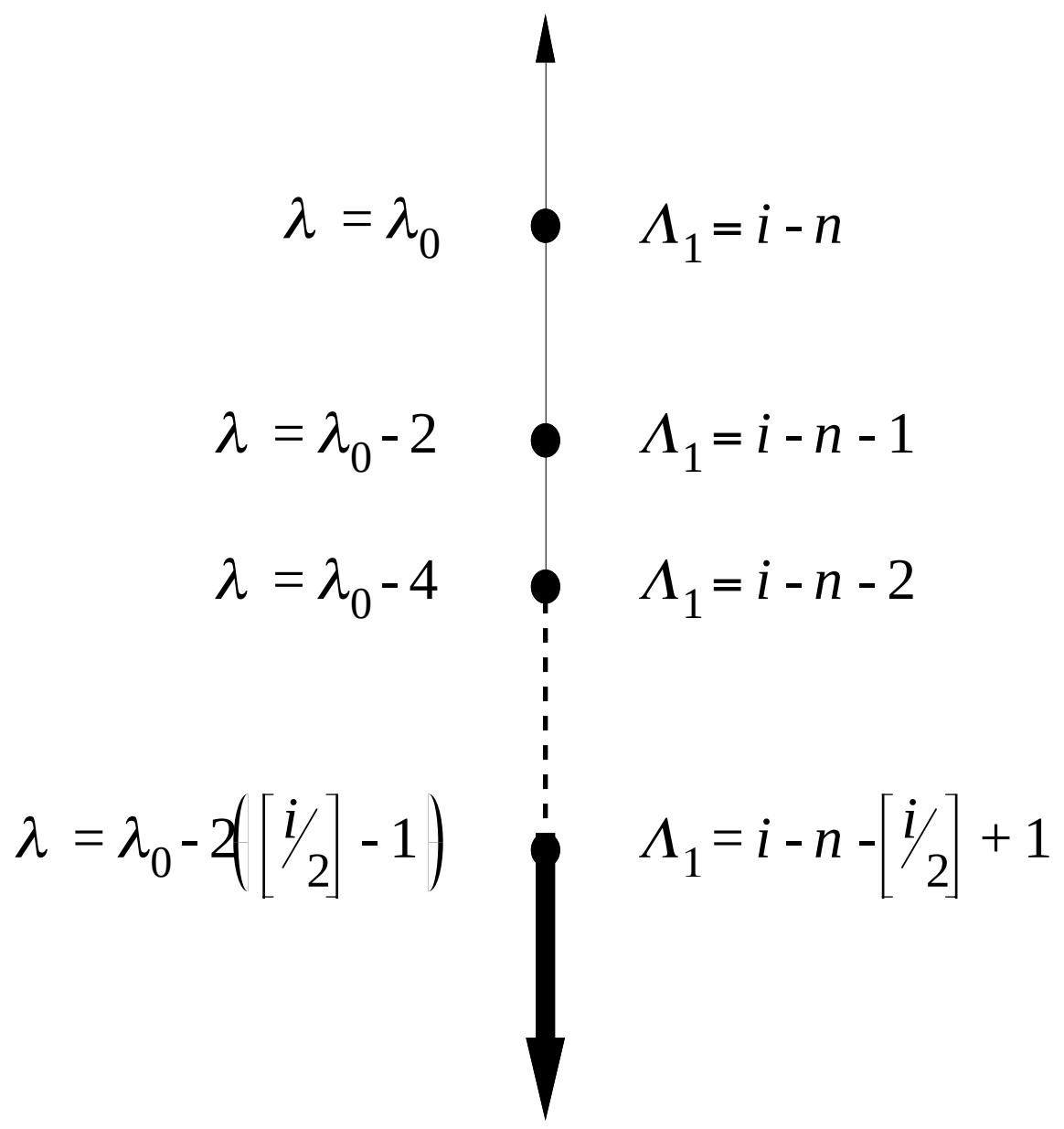

**Example so*(16)**

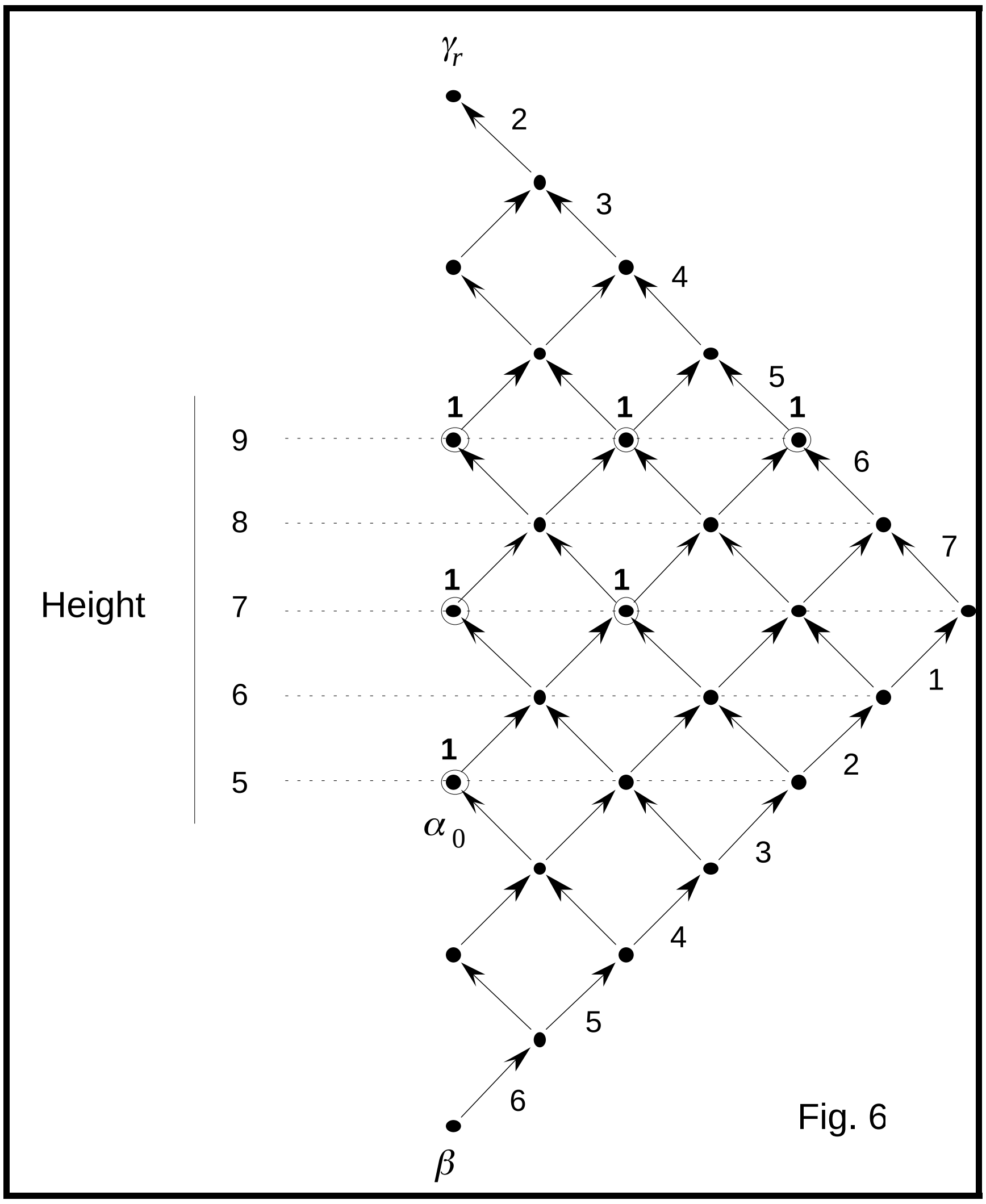

Fig. 6

We consider, the following conditions on $\Lambda_0$ :

$\bullet\Lambda_0, \mu_1 \circledR = 0$ for $1 \leq i \leq 5$

$\bullet\Lambda_0, \mu_6 \circledR = n_6 > 1$ $\bullet\Lambda_0, \mu_7 \circledR = n_7$

Then

$$\alpha_0 = \beta + 2\mu_6 + \mu_7 + \mu_5$$

with height 5

The condition $\bullet\Lambda + R, \alpha_0 \circledR = 1$ implies

$$(\Lambda_0, \alpha_0) + \lambda_0 + 5 = 1 \quad ; \quad \lambda_0 = -4$$

For $\lambda_q = \lambda_0 + \lambda_s = -6$ we have $\Lambda' = \Lambda_0 - 6\varepsilon$ and

$$\bullet\Lambda + R, \alpha^1{}_6 \circledR = \lambda_q + 6 = 0$$

then it is not a valid root. However for the roots the height of which is 7 we have

$$\bullet\Lambda + R, \alpha^1{}_7 \circledR = \bullet\Lambda + R, \alpha^2{}_7 \circledR = 1$$

and there will be a second order polynomial which will be missing with highest weight:

$$\Lambda_0 - 6\varepsilon + R - \alpha^1{}_7 - \alpha^2{}_7$$

For $\lambda_q = \lambda_0 + 2\lambda_s = -8$ , the roots are those with height 9 :

$$\bullet\Lambda + R, \alpha^1{}_9 \circledR = \bullet\Lambda + R, \alpha^2{}_9 \circledR = \bullet\Lambda + R, \alpha^3{}_9 \circledR = \lambda_q + 9 = 1$$

and a third order polynomial will be missing with highest weight

$$\Lambda_0 - 8\varepsilon + R - \alpha^1{}_9 - \alpha^2{}_9 - \alpha^3{}_9$$

For $\lambda_q < -8$ there is no roots for which there exists $k_1$-dominance then the first possible place for non unitarity is $\lambda_q = -8$ .

The highest weight vectors are in this case, formally, the followings

**1)** Height 5

$$E_{-\mu_7}^{-n_7-1}\, E_{-\mu_6}^{-n_6-n_7-1}\, E_{-\mu_5}^{-n_6-n_7-2}\, E_{-\mu_7}^{-n_6}\, E_{-\mu_6}^{-n_6-1}\, E_{-\beta}$$

$$\times\ E_{-\mu_6}^{n_6+2}\, E_{-\mu_7}^{n_6+1}\, E_{-\mu_5}^{n_6+n_7+3}\, E_{-\mu_6}^{n_6+n_7+2}\, E_{-\mu_7}^{n_7+1}$$

with $\Lambda = \Lambda_0 - 4\varepsilon + R - \beta\ -\mu_5 - 2\mu_6 - \mu_7$

**2)** Height 7

$$E_{-\mu_4}^{-1}\, E_{-\mu_7}^{-n_7-1}\, E_{-\mu_6}^{-n_6-n_7-1}\, E_{-\mu_5}^{-n_6-n_7-3}\, E_{-\mu_7}^{-n_6}\, E_{-\mu_6}^{-n_6-2}\, E_{-\beta}^{2}$$

$$\times\ E_{-\mu_6}^{n_6+4}\, E_{-\mu_7}^{n_6+2}\, E_{-\mu_5}^{n_6+n_7+5}\, E_{-\mu_6}^{n_6+n_7+3}\, E_{-\mu_7}^{n_7+1}\, E_{-\mu_4}^{3}\, E_{-\mu_3}\, E_{-\mu_5}$$

with $\Lambda = \Lambda_0 - 6\varepsilon + R - 2\beta\ -\mu_3 - 2\mu_4 - 3\mu_5 - 4\mu_6 - 2\mu_7$

**3)** Height 9

$$E_{-\mu_3}^{-1}\, E_{-\mu_4}^{-2}\, E_{-\mu_7}^{-n_7-1}\, E_{-\mu_6}^{-n_6-n_7-1}\, E_{-\mu_5}^{-n_6-n_7-4}\, E_{-\mu_7}^{-n_6}\, E_{-\mu_6}^{-n_6-3}\, E_{-\beta}^{3}$$

$$\times\ E_{-\mu_6}^{n_6+6}\, E_{-\mu_7}^{n_6+3}\, E_{-\mu_5}^{n_6+n_7+7}\, E_{-\mu_6}^{n_6+n_7+4}\, E_{-\mu_7}^{n_7+1}\, E_{-\mu_4}^{5}\, E_{-\mu_3}^{4}\, E_{-\mu_5}^{2}\, E_{-\mu_2}^{2} E_{-\mu_1}\, E_{-\mu_4}$$

with $\Lambda = \Lambda_0 - 8\varepsilon + R - 3\beta\ -\mu_1 - 2\mu_2 - 3\mu_3 - 4\mu_4 - 5\mu_5 - 6\mu_6 - 3\mu_7$

## $e_6$

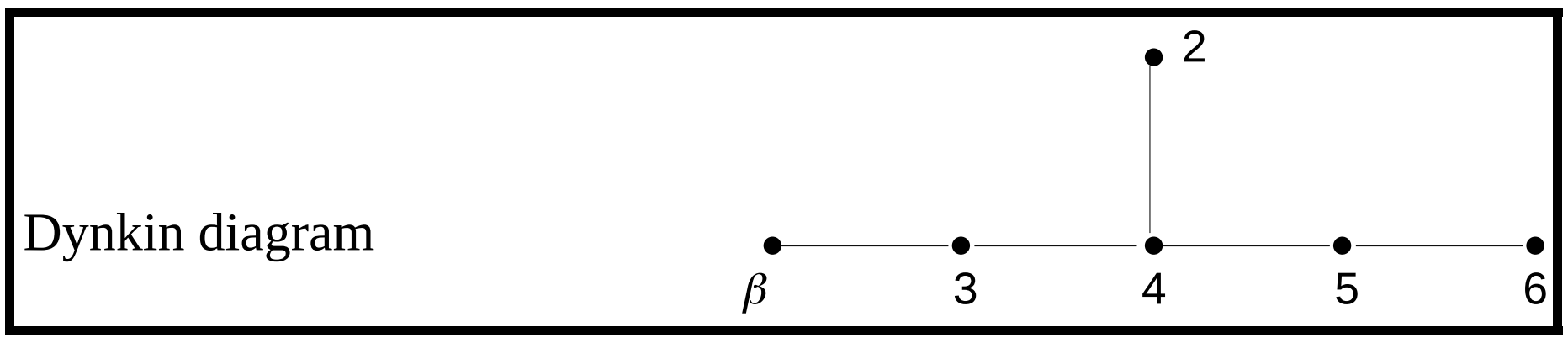

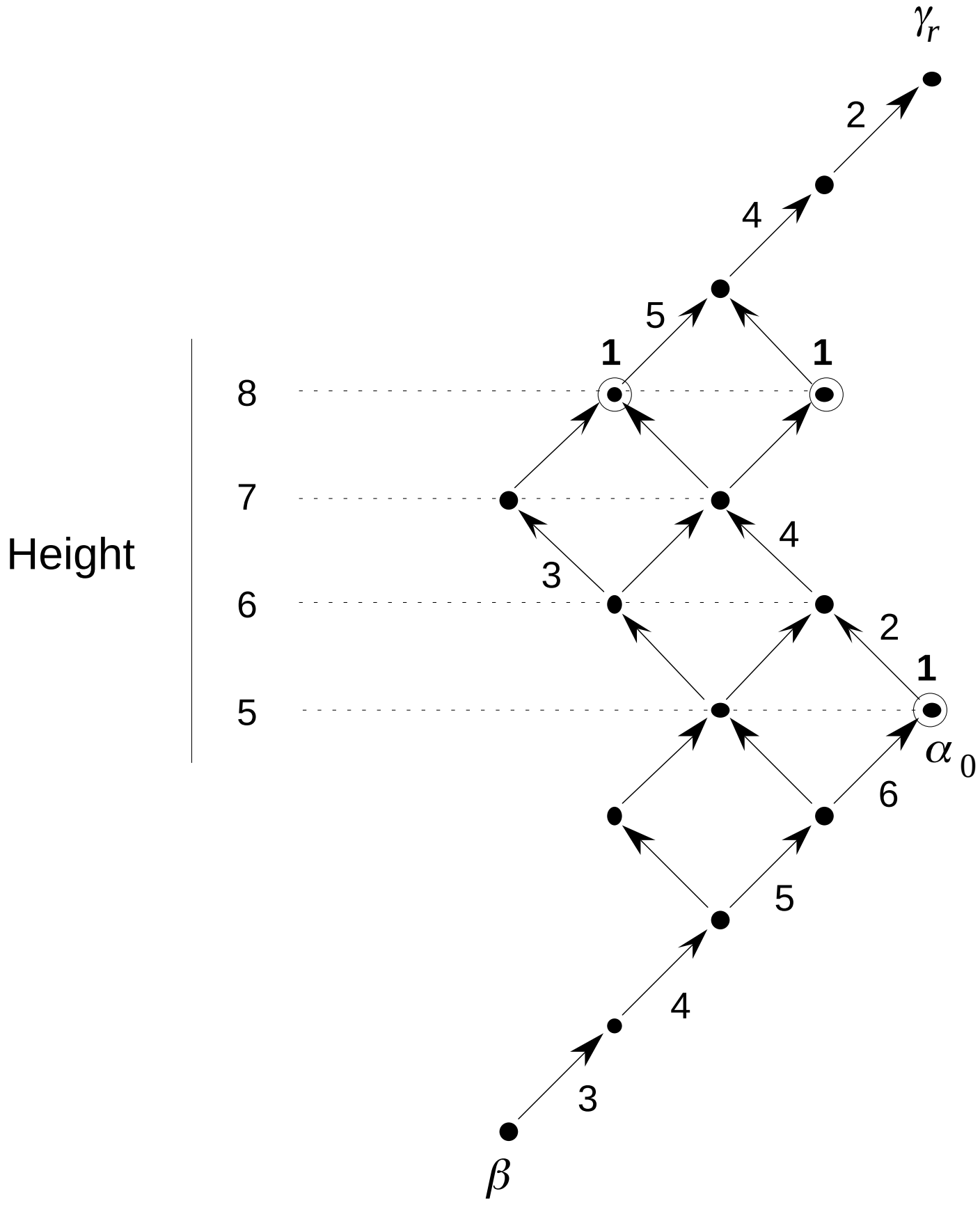

Fig. 7

Here $\Lambda = \lambda_0 + \lambda\varepsilon$ with $\varepsilon = \left(0, 0, 0, 0, 0, -\tfrac{2}{3}, -\tfrac{2}{3}, -\tfrac{2}{3}\right)$ For $\Lambda_0 \neq 0$ the only case for which the reduction level is strictly higher than one (all the rest are excluded for $k_1$-dominance arguments ) is the following

$$\langle \Lambda_0, \mu_i \rangle = 0 \qquad 2 \leq i \leq 5$$

$$\langle \Lambda_0, \mu_6 \rangle = n > 0$$

Applying Jakobsen method we obtain

$$\alpha_0 = \beta + \mu_3 + \mu_4 + \mu_5 + \mu_6$$

with height 5. From the condition $\langle \Lambda + R, \alpha_0 \rangle = 1$ we obtain

$$1 = \langle \Lambda, \alpha_0 \rangle + \langle R, \alpha_0 \rangle = \langle \Lambda, \gamma_r \rangle + 5 = \lambda_0 + 5\,; \quad \lambda_0 = -4$$

For $\lambda_q = \lambda_0 + \lambda_s = -7$

$$\langle \Lambda + R, \alpha^1{}_8 \rangle = \langle \Lambda + R, \alpha^2{}_8 \rangle = 1$$

then a second order polynomial will be missing with highest weight:

$$\Lambda_0 - 7\varepsilon + R - \alpha^1{}_8 - \alpha^2{}_8$$

The value $\lambda_q = -7$ is the first possible place for non unitarity.

The highest weight vector corresponding to the highest weight

$\Lambda_0 - 4\varepsilon + R - \beta - \mu_3 - \mu_4 - \mu_5 - \mu_6$ is, formally:

$$\boxed{E^{-n}_{-\mu_6}\, E^{-n-1}_{-\mu_5}\, E^{-n-2}_{-\mu_4}\, E^{-n-3}_{-\mu_3}\, E_{-\beta}\, E^{n+4}_{-\mu_3}\, E^{n+3}_{-\mu_4}\, E^{n+2}_{-\mu_5}\, E^{n+1}_{-\mu_6}}$$

## $e_7$

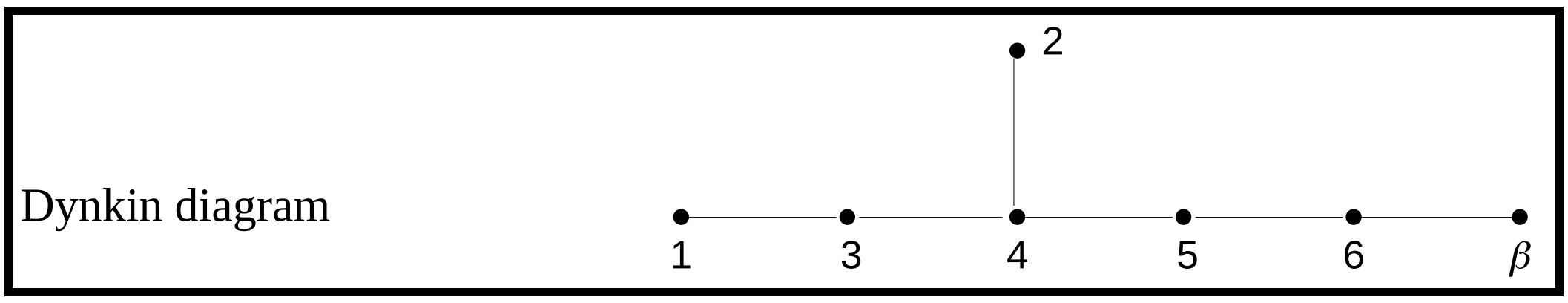

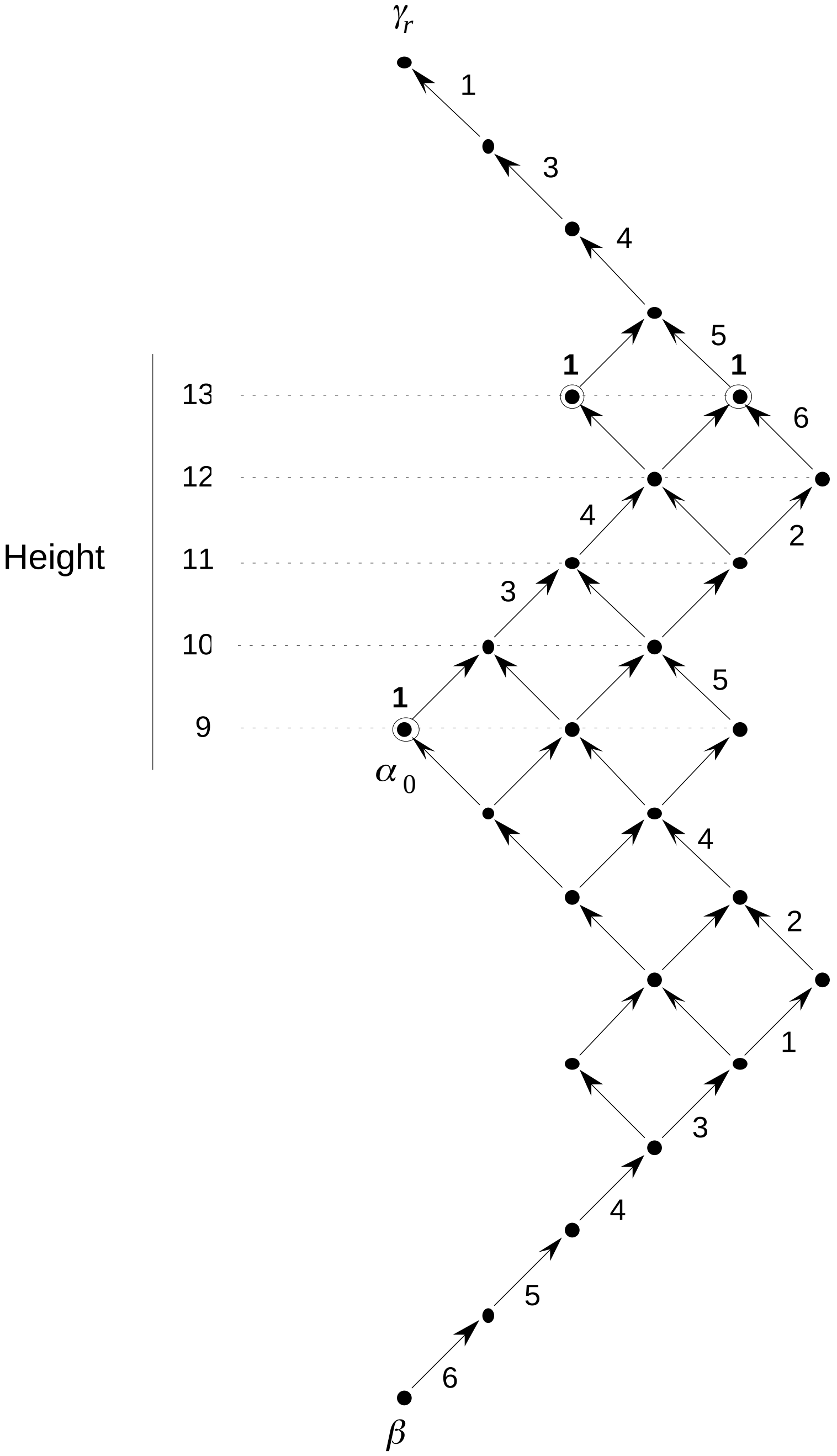

Fig. 8

For this case $\varepsilon = \left(0, 0, 0, 0, 0, 1, -\tfrac{1}{2}, \tfrac{1}{2}\right)$. As in $e_6$ we must only consider one case:

$$\bullet\Lambda_0, \mu_i ® = 0 \qquad 1 \le i \le 5$$

$$\bullet\Lambda_0, \mu_6 ® = n > 0$$

with those conditions

$$\alpha_0 = \beta + 2\mu_6 + 2\mu_5 + 2\mu_4 + \mu_3 + \mu_2$$

with height 9, then

$$1 = \bullet\Lambda, \alpha_0 ® + \bullet R, \alpha_0 ® = \bullet\Lambda, \alpha_0 ® + 9 = \lambda_0 + 9 \; ; \quad \lambda_0 = -8$$

For $\lambda_q = -12$

$$\bullet\Lambda + R, \alpha^1{}_{13} ® = \bullet\Lambda + R, \alpha^2{}_{13} ® = 1$$

and we obtain, for $\lambda_q = -12$ , a missing second order polynomial with highest weight

$$\Lambda_0 - 12\varepsilon + R - \alpha^1{}_{13} - \alpha^2{}_{13}$$

The highest weight vector in height 9 is, formally:

$$\mathrm{E}_{-\mu_2}^{-1} \mathrm{E}_{-\mu_3}^{-1} \mathrm{E}_{-\mu_4}^{-3} \mathrm{E}_{-\mu_5}^{-4} \mathrm{E}_{-\mu_6}^{-n-4} \mathrm{E}_{-\mu_2}^{-2} \mathrm{E}_{-\mu_3}^{-2} \mathrm{E}_{-\mu_4}^{-5} \mathrm{E}_{-\mu_5}^{-n-5} \mathrm{E}_{-\mu_6}^{-1} \mathrm{E}_{-\mu_2}^{-3} \mathrm{E}_{-\mu_3}^{-3} \mathrm{E}_{-\mu_4}^{-n-6} \mathrm{E}_{-\mu_5}^{-2} \mathrm{E}_{-\mu_6}^{-1}$$

$$\times\ \mathrm{E}_{-\mu_2}^{-n-3} \mathrm{E}_{-\mu_3}^{-n-3} \mathrm{E}_{-\mu_4}^{-n-2} \mathrm{E}_{-\mu_5}^{-n-1} \mathrm{E}_{-\mu_6}^{-n} \mathrm{E}_{-\beta} \mathrm{E}_{-\mu_6}^{n+1} \mathrm{E}_{-\mu_5}^{n+2} \mathrm{E}_{-\mu_4}^{n+3} \mathrm{E}_{-\mu_3}^{n+4} \mathrm{E}_{-\mu_2}^{n+4}$$

$$\times\ \mathrm{E}_{-\mu_6} \mathrm{E}_{-\mu_5}^{2} \mathrm{E}_{-\mu_4}^{n+7} \mathrm{E}_{-\mu_3}^{3} \mathrm{E}_{-\mu_2}^{3} \mathrm{E}_{-\mu_6} \mathrm{E}_{-\mu_5}^{n+6} \mathrm{E}_{-\mu_4}^{5} \mathrm{E}_{-\mu_3}^{2} \mathrm{E}_{-\mu_2}^{2} \mathrm{E}_{-\mu_6}^{n+5} \mathrm{E}_{-\mu_5}^{4} \mathrm{E}_{-\mu_4}^{3} \mathrm{E}_{-\mu_3} \mathrm{E}_{-\mu_2}$$

with highest weight

$$\Lambda_0 - 8\varepsilon + R - \beta - \mu_2 - \mu_3 - 2\mu_4 - 2\mu_5 - 2\mu_6$$